\documentclass[twocolumn,superscriptaddress]{revtex4-2}

\usepackage{prl_style}

\newcommand{\lammod}{\lambda_{\mathrm{mod}}}
\newcommand{\lamrec}{\lambda_{\text{rec}}}

\begin{document}



\title{\papertitle}

\author{H.E.~Read}
\affiliation{School of Engineering and Applied Sciences, Harvard University, Cambridge,	Massachusetts 02138, USA}

\author{G.~Risso}
\affiliation{School of Engineering and Applied Sciences, Harvard University, Cambridge,	Massachusetts 02138, USA}

\author{A.~Djellouli}
\affiliation{School of Engineering and Applied Sciences, Harvard University, Cambridge,	Massachusetts 02138, USA}

\author{K.~Bertoldi}
\affiliation{School of Engineering and Applied Sciences, Harvard University, Cambridge,	Massachusetts 02138, USA}

\author{A.~Lazarus}
\affiliation{Department of Mathematics, Massachusetts Institute of Technology, Cambridge, Massachusetts 02139, USA}
\affiliation{Institut Jean Le Rond d'Alembert, CNRS UMR7190, Sorbonne Universit\'e, Paris, France}

\thanks{Corresponding authors: bertoldi@seas.harvard.edu and arnaud.lazarus@upmc.fr}

\begin{abstract}
Parametric instabilities are a known feature of periodically driven dynamic systems; at particular frequencies and amplitudes of the driving modulation, the system's quasi-periodic response undergoes a frequency lock-in, leading to a periodically unstable response.
Here, we demonstrate an analogous phenomenon in a purely static context. We show that the buckling patterns of an elastic beam resting on a modulated Winkler foundation display the same kind of frequency lock-in observed in dynamic systems.
Through simulations and experiments, we reveal that compressed elastic strips with modulated height alternate between predictable quasi-periodic and periodic buckling modes. Our findings uncover previously unexplored analogies between structural and dynamic instabilities, highlighting how even simple elastic structures can give rise to rich and intriguing behaviors.

\end{abstract}


\maketitle

\begin{figure}[hb]
    \centering 
    \includegraphics{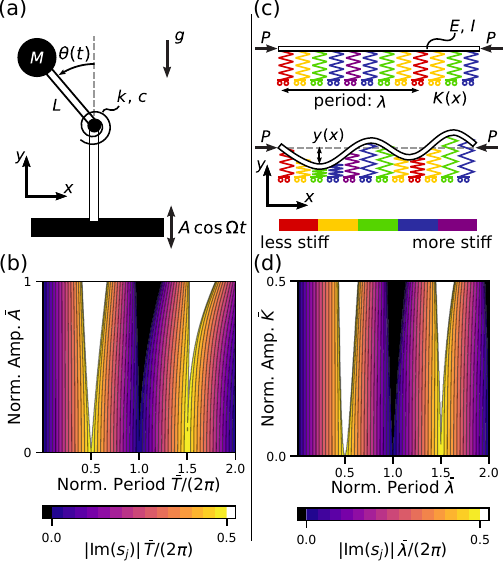}
    \caption{\textbf{Frequency lock-in in dynamics and statics} (a) An inverted parametric pendulum. (b) Maximum normalized imaginary component of Floquet exponents $s_j$ for the inverted pendulum in the first Brillouin zone. (c) A modulated Winkler foundation. (d) Maximum normalized imaginary component of $s_j$ for the modulated Winkler foundation.}
    \label{fig:main_fig1}
\end{figure}

\textit{Introduction}---
Parametric instabilities are a well studied class of instability  that arise in systems subjected to periodic driving. These phenomena appear across a wide range of physical systems, including 
pendula \cite{sanmartin1984botafumeiro, grandi2021enhancing}, surface gravity waves \cite{faraday1831xvii, benjamin1954stability},
and Floquet time crystals \cite{heugel2019classical, zhang2017observation}.
In all these systems, specific combinations of excitation frequency and amplitude cause perturbations around equilibrium to become dynamically amplified---a phenomenon known as frequency lock-in. A classic example of a system exhibiting frequency lock-in is the inverted pendulum \cite{smith1992experimental,acheson1993upside}(Fig.~\ref{fig:main_fig1}a). 
When its base is oscillated vertically, it is well known that a statically unstable upright position can become stable for specific combinations of oscillation amplitude and frequency. 
More generally, this Floquet system exhibits alternating stable and unstable regions in the amplitude–frequency space \cite{grandi2023new}, with the unstable regions forming characteristic tongue-like shapes.
These so-called instability tongues correspond to parametric resonances, where the quasi-periodic response ``locks in" to become periodic, with a period either twice or equal to the driving period $\bar{T}$ (white and black regions in Fig.~\ref{fig:main_fig1}b, respectively), while growing exponentially in time to eventually reach a limit cycle. 

Another extensively studied class of instabilities is structural instabilities, such as buckling, creasing, and snapping \cite{audoly2000elasticity}.
These instabilities arise in elastic structures when compressive forces exceed a critical threshold \cite{bazant2010stability}, resulting in sudden and often dramatic shape changes. 
These phenomena have attracted considerable attention in recent years. 
They have been harnessed to enable functional behavior such as reconfigurable, adaptive, and extreme shape morphing \cite{krieger2012extreme, reis2015perspective, shim2012buckling, lazarus2015soft, RISSO2021113946, yang2024complex, lee2014tilted, doare2011piezoelectric, michelin2013energy}, and have served as powerful tools for probing complex physical phenomena, offering intuitive platforms to visualize effects that are otherwise difficult to observe. 
Examples include the amplification of chirality \cite{kang2013buckling}, as well as domain wall motion \cite{deng2020characterization}, and phase transitions \cite{jin2020guided, nadkarni2016unidirectional}. 
In addition, researchers have begun to draw analogies between structural and dynamic instabilities.
For instance, period-doubling bifurcations, a hallmark of nonlinear dynamic systems, have been observed in the post-buckling regime of compressed elastic structures \cite{brau2011multiple,brau2013wrinkle}, highlighting the rich and complex behavior of elastic systems.

Motivated by these efforts, we show that the buckling behavior of certain elastic structures can display key features typically associated with parametric resonance, most notably the emergence of frequency lock-in phenomena. To explore this analogy between parametric instabilities and structural buckling, we begin by analyzing an axially compressed beam resting on a bed of springs with spatially modulated stiffness (Fig.~\ref{fig:main_fig1}c). 
Buckling occurs when the applied force exceeds a critical threshold, giving rise to a wavy deformation pattern corresponding to the system’s critical mode. 
Remarkably, in close analogy to parametric instabilities, the resulting buckling modes alternate between periodic and quasi-periodic behavior depending on the amplitude and wavelength of the stiffness modulation with the periodic regions forming distinctive tongue-like shapes (Fig.~\ref{fig:main_fig1}d). 
Building on these results, we construct a three-dimensional analogue consisting of a thin elastic strip with periodically varying height, clamped and compressed along its bottom edge. 
Through a combination of experiments and numerical simulations, we show that this system reproduces the essential behavior of the spring-supported beam and similarly exhibits an alternation between periodic and quasi-periodic buckling patterns that is fully governed by the geometric modulation of the strip.

\textit{A boundary value analogue}---We start by considering an elastic beam with bending modulus $EI$ on a  linear Winkler foundation with a periodic stiffness per unit length (see Fig.~\ref{fig:main_fig1}c).
\begin{equation}
    K(x) = K_0 + K_1 \cos{\frac{2\pi}{\lambda} x},
\end{equation}
where $K_0$ is the base stiffness, $K_1$ is the amplitude of modulation, and $\lambda$ is the wavelength of modulation.
The linearized equilibrium equation governing its transverse deflection $y(x)$ about $y=0$ upon application of an axial load $P$ takes the form \cite{brau2013wrinkle,lagrange2012solution}

\begin{equation}\label{eqn:winkler_nondim_main}
    y^{(4)} + \bar{P} y'' + 16 \pi^4 \left(1 + \bar{K} \cos{\frac{2\pi}{\bar{\lambda}} \bar{x}} \right) \, y = 0,
\end{equation}
where $\bar{K} = K_1 / K_0$ and $\bar{P} = 4\pi^2 P / \sqrt{K_0 EI}$. Moreover, $\bar{\lambda} = \lambda / \lambda_0$ and  $\bar{x} = x / \lambda_0$, with 

\begin{equation}\label{l0}
    \lambda_0 = 2\pi \left(\frac{EI}{K_0} \right)^{1/4},
\end{equation}
which corresponds to the buckling wavelength of an inextensible Euler–Bernoulli beam resting on an unmodulated Winkler foundation  (i.e., $K_1 = 0$) \cite{brau2013wrinkle}.
(See Supplementary information Section II for details). 
Eq.~(\ref{eqn:winkler_nondim_main}) shows that, much like the inverted pendulum, the response of the modulated Winkler foundation is governed by a linear differential equation with periodically varying coefficients. 
According to Floquet-Bloch theory \cite{richards2012analysis}, the solutions to such equations can be expressed in the reciprocal space as 

\begin{equation}\label{EqFLoquet}
  \hat{y}(\bar{x}) = \sum_{j = 1}^4 \sum_{h \in \mathbb{Z}} r_j^h e^{(ih2\pi/\bar{\lambda} + s_j) \bar{x}},
\end{equation}
where $s_j \in \mathbb{C}$ are the Floquet exponents and $r_j^h \in \mathbb{C}$ are harmonics of a periodic funtion and are determined by boundary conditions.
(see Supporting Information Section III for details).
Instead of computing the full solution, we focus on determining the Floquet exponents within the first Brillouin zone, $|\Im(s_j)| \leq \pi / \bar{\lambda}$, using the Hill (spectral) method \cite{hill1886part, deconinck2007spectruw, bentvelsen2017modal}, as they offer key insights into the linear behavior of the system.
A nontrivial bounded solution $y(\bar{x}) \neq 0$ exists whenever at least one Floquet exponent has zero real part (i.e., $\Re(s_j) = 0$). 
Therefore, we define the critical buckling load $P_{cr}$ as the minimum applied load at which the real part of at least one Floquet exponent vanishes.  
The imaginary parts of the Floquet exponents satisfying $\Re(s_j) = 0$ at $P_{cr}$ reveal the spectral content of the emerging buckling mode.
From Eq.~(\ref{EqFLoquet}), it follows that the wavenumber spectrum of $y(\bar{x})$ is given by $\Im(s_j) + h 2\pi/\bar{\lambda}$, where $h \in \mathbb{Z}$.
If $\Im(s_j) = 0$, the spectrum reduces to integer multiples of $2\pi/\bar{\lambda}$, meaning the buckling mode is $\bar{\lambda}$-periodic.  
If $|\Im(s_j)| = \pi/\bar{\lambda}$, the spectrum instead reduces to half-integer multiples of $2\pi/\bar{\lambda}$, so the mode becomes $2 \bar{\lambda}$-periodic.  
In all other cases, the buckling mode is quasi-periodic, characterized by two incommensurate wavenumbers.

In Fig.~\ref{fig:main_fig1}d we show the evolution of the maximum $\Im(s_j) \bar{\lambda} / (2\pi)$ in the first Brillouin zone as a function of $\bar{\lambda}$ and $\bar{K}$ at the onset of buckling. 
We again see the emergence of ``frequency'' lock-in tongues; in these regions, the wavenumber response ``locks-in'' to either double (white) or equal (black) to the ``driving'' wavelength $\bar{\lambda}$ of the foundation.
As with the inverted pendulum, outside of these  tongues, our solution is quasi-periodic.

\begin{figure}[ht]
    \centering
    \includegraphics{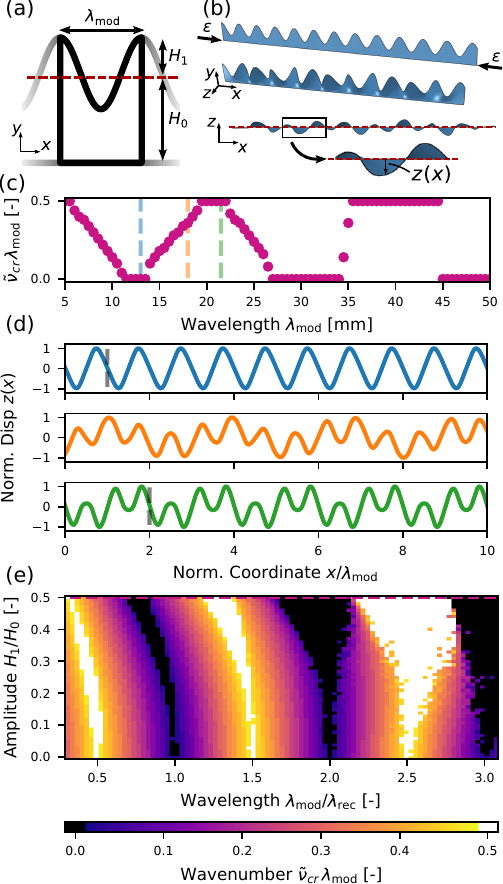}
    \caption{\textbf{Wavenumber response of the compressed elastic strip} (a) Schematic of our system. (b) When compressed along its bottom edge, the strip buckles out of plane. (c) Critical wavenumbers as a function of $\lammod$ at $H_1 / H_0 = 0.5$. (d) Some example reconstructions; periodicity marked with grey dashed line. (e) Wavenumber lock-in tongues; the horizontal slice that generates 2c is shown in magenta.}
    \label{fig:main_fig2}
\end{figure}

\textit{A physical realization---}In Fig.~\ref{fig:main_fig1}, we show that the buckling behavior of an elastic beam resting on a Winkler foundation with periodically modulated stiffness exhibits wavenumber lock-in tongues.  
Motivated by this result, we design a physical platform that reproduces the same effect. 
We know that an axially compressed rectangular strip of height $H_0$ buckles out of plane, forming a wavy pattern with wavelength ~\cite{mora2006buckling,Siefert2021stretch}

\begin{equation}\label{l0A}
    \lamrec = 3.256 \, H_0.
\end{equation}
Comparing Eq.~(\ref{l0}) and Eq.~(\ref{l0A}), we see that the height plays the role of an effective stiffness for the out-of-plane motion of the strip; by periodically modulating the height as a function of $x$, we expect a buckling behavior analogous to the Winkler foundation of Figs.~\ref{fig:main_fig1}c-d.
In practice, we introduce a periodic modulation of the height

\begin{equation}
    H(x) = H_0 + H_1 \cos{\frac{2\pi}{\lammod} x},
\end{equation}
where $H_1$ is the amplitude of modulation, and $\lammod$ is the ``driving" wavelength (Fig.~\ref{fig:main_fig2}a).
Because the strip buckles out of plane, we employ Bloch wave analysis \cite{brillouin1946wave}, a 3D generalization of the Floquet analysis used for the Winkler foundation.
We implement this analysis within a Finite Element (FE) framework to determine the buckling load and examine the characteristics of the emerging buckling modes. 

Specifically, we consider a unit cell and first compress it to some axial strain $\epsilon$ under periodic boundary conditions. We then investigate Bloch-type small perturbations superimposed on the axially compressed finite state of deformation 

\begin{equation}
    \mathbf{u}^{\text{right}} = \mathbf{u}^{\text{left}} \, e^{i 2 \pi \tilde{\nu} \lammod},
\end{equation}
where $\tilde{\nu}$ is a wavenumber and $\mathbf{u}^{\text{right}}$ and $\mathbf{u}^{\text{left}}$ denote the displacement of superimposed perturbation applied to the right and left edges, respectively. 
Buckling is triggered at the lowest applied strain $\epsilon$ for which a vanishing eigenfrequency exists  for wavenumbers $\tilde{\nu}$ within the first Brillouin zone, i.e. $\tilde{\nu} \in [0, 1/\lammod)$; due to symmetry in our system, we can further restrict ourselves to $\tilde{\nu} \in [0, 0.5/\lammod]$
 (see Supplementary Information Section IV.C for details).
In Fig.\ref{fig:main_fig2}c, we report the calculated critical values of $\tilde{\nu}$, denoted as $\tilde{\nu}_{cr}$, as a function of $\lammod$ for a strip with $H_1 / H_0 = 0.5$. 
As in the case of the Winkler foundation, we observe locked-in regions at $\tilde{\nu}_{cr} \lammod = 0$ and $0.5$, corresponding to periodic solutions with wavelengths $\lammod$ and $2 \lammod$, respectively. 
This is confirmed by the reconstructed buckling modes, shown in Fig.~\ref{fig:main_fig2}d; in analogy to the transverse displacement $y(x)$ in the Winkler foundation, we display the out-of-plane displacement along the top edge of the strip, $z(x)=u_z^{\text{top}}$ (Fig.~\ref{fig:main_fig2}b).

Finally, we compute the critical wavenumbers $\tilde{\nu}_{cr}$ for 51 amplitude values in the range $H_1 / H_0 \in [0, 0.5]$ and 91 modulation wavelengths $\lammod \in [5, 50]$ (corresponding to $\lammod / \lamrec \in [0.307, 3.07]$), and present the results in Fig.~\ref{fig:main_fig2}e.  
As in the case of the modulated Winkler foundation, we observe the emergence of ``instability tongues,'' where wavenumber lock-in occurs and the solution becomes periodic, confirming that an axially compressed elastic strip with modulated height  displays features typically associated with parametric resonance.

Guided by the results of Fig.~\ref{fig:main_fig2}, we fabricate four samples with $H_1 / H_0 = 0.5$ and modulation wavelengths $\lammod = 8.2$, 9.4, 9.8, 10.2, and 12.2 mm.  
Based on the Bloch wave analysis, we expect the sample with $\lammod = 12.2$~mm to exhibit a periodic postbuckling deformation with period $\lammod$, while all other samples should display quasi-periodic postbuckling deformation.  
The samples are cast from an elastomeric material (Zhermack Elite Double 32) using a molding process (see Supplementary Information Section V for details). All strips have a length $L = 200$ mm, a nominal height $H_0 = 5$ mm, and thickness $t = 0.5$ mm.

\begin{figure}[ht]
    \centering
    \includegraphics{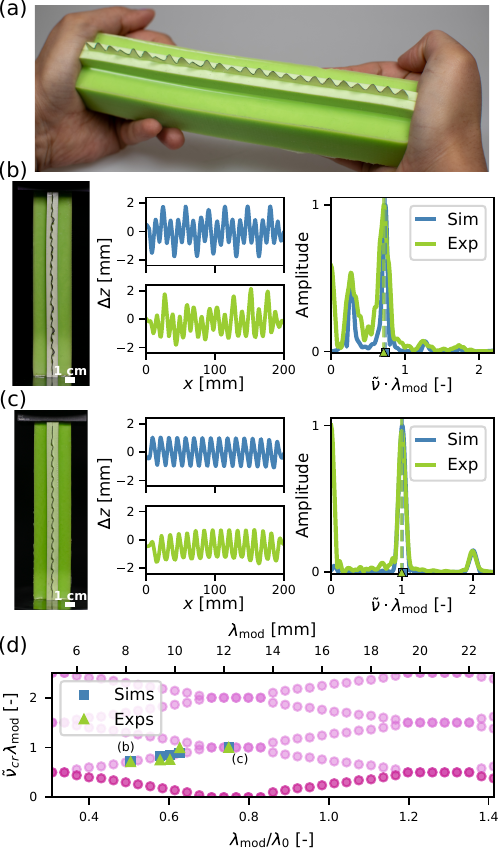}
    \caption{\textbf{Experimental analysis} (a) An experimental sample with the top edge highlighted with a black marker. (b, c) Experimental, simulation deformation and Fourier transform for (b) $\lammod = 8.2$ mm and $H_1 / H_0 = 0.5$ and (c) $\lammod = 12.2$ mm and $H_1 / H_0 = 0.5$. (d) Peak wavenumber location in post-buckling simulation vs. experimental samples; Bloch wave simulation data is shown in pink.}
    \label{fig:main_fig3}
\end{figure}


To apply axial compression, each thin strip is bonded to a larger block of the same material (Fig.~\ref{fig:main_fig3}a), which is then compressed to induce buckling in the strip.
We mark the top edge of each strip in black and track its deformation during compression. The resulting buckling patterns for $\lammod = 8.2$ mm and $\lammod = 12.2$ mm are shown in Fig.~\ref{fig:main_fig3}b and Fig.~\ref{fig:main_fig3}c, respectively. 
All samples were compressed to between $5$--$6.5$\% strain, which is much larger than the compressive strain needed for buckling onset 
(see Supplementary Information Sections VI for details).
As expected, the sample with $\lammod = 8.2$ mm exhibits a disordered postbuckling deformation, while the sample with $\lammod = 12.2$ mm shows an ordered, apparently periodic postbuckling pattern.
To quantify these observations, we perform a discrete Fourier transform (DFT) on the experimental data. For the sample with $\lammod = 12.2$ mm, the DFT shows a spectral peak at $\tilde{\nu} \lammod= 1.0$, confirming that the deformation is  periodic with spatial period equal to $\lammod$.  
In contrast, the spectrum for $\lammod = 8.2$ mm exhibits a peak at $\tilde{\nu} \lammod = 0.71$, indicating that the response is quasi-periodic.

Concurrently, we perform FE simulations on strips of the same dimensions used in the experiments ($L = 200$ mm).  
As in the experiments, the models are compressed into the post-buckling regime, with the first linear buckling mode introduced as a geometric imperfection to initiate the instability (see Supplementary Information Section IV.B.2 for details).
As shown  in Fig.~\ref{fig:main_fig3}b and Fig.~\ref{fig:main_fig3}c, we observe good agreement between simulations and experiments, both qualitatively and quantitatively, in the out-of-plane deformation of the top edge.

Finally, we compare these results with those obtained from the Bloch wave analysis, theoretically valid for an infinite strip length.  
In Fig.~\ref{fig:main_fig3}d, the magenta markers indicate the values of $\tilde{\nu}_{cr} \lammod$ predicted by the Bloch analysis (same data as in Fig.~\ref{fig:main_fig2}c, but here we also display the extended region in a lighter shade), while  the blue and green markers represent the dominant DFT peaks extracted from the numerical simulations and experimental data, respectively. 
We observe good overall agreement between the experimental and numerical results, demonstrating that the transition from quasi-periodic to periodic behavior is accurately captured even for the relatively short strip length used in our study.  
Importantly, this agreement is not limited to the two representative cases shown in Fig.~\ref{fig:main_fig2}b and Fig.~\ref{fig:main_fig2}c, but also holds across the three additional tested samples (see Supplementary Information Section VI).

In summary, we have demonstrated the realization of a static analog of parametric instabilities. We first established this in the archetypal case of a beam on a modulated Winkler foundation, where Floquet analysis revealed the same frequency lock-in tongues that characterize dynamic systems exhibiting parametric instabilities. We then extended the concept to a physical platform: an elastic strip with modulated thickness. Using a combination of Bloch-wave analysis and experiments, we again observed the characteristic lock-in tongues, confirming the predictable alternation between quasi-periodic and fully periodic buckling patterns.

The wavenumber lock-in phenomenon uncovered in this paper in the framework of elastic buckling should offer promising opportunities, for example, in the spectral enrichment of elastic wrinkling patterns \cite{chen2004herringbone,lee2014tilted, lee2017mechano, wu2022initiation}. For the wrinkling pattern modulation to be real-time, one could control the underlying periodic geometry with programmable and active elastic meta-materials \cite{lowe2005dielectric, florijn2014programmable, ghatak2020observation}. 
Another promising direction is to extend the analogy with parametric instabilities, where recent work has shown that, in a certain modulation regime, Floquet-Bloch theory becomes mathematically equivalent to the linear algebra of quantum mechanics \cite{lazarus2025optimal,lazarus2025meaningful}---opening new opportunities for exotic functionalities in buckled elastic systems.



\section*{Acknowledgments} \label{sec:acknowledgements}
This work was supported by the Simons Collaboration on Extreme Wave Phenomena Based on
Symmetries. G.R. acknowledges support from the Swiss National Science Foundation under Grant No. P500PT-217901. A.L acknowledges support from the French National Center for Scientific Research for hosting him in a CNRS delegation.

\bibliography{Bibliobus}

\begin{thebibliography}{47}%
\makeatletter
\providecommand \@ifxundefined [1]{%
 \@ifx{#1\undefined}
}%
\providecommand \@ifnum [1]{%
 \ifnum #1\expandafter \@firstoftwo
 \else \expandafter \@secondoftwo
 \fi
}%
\providecommand \@ifx [1]{%
 \ifx #1\expandafter \@firstoftwo
 \else \expandafter \@secondoftwo
 \fi
}%
\providecommand \natexlab [1]{#1}%
\providecommand \enquote  [1]{``#1''}%
\providecommand \bibnamefont  [1]{#1}%
\providecommand \bibfnamefont [1]{#1}%
\providecommand \citenamefont [1]{#1}%
\providecommand \href@noop [0]{\@secondoftwo}%
\providecommand \href [0]{\begingroup \@sanitize@url \@href}%
\providecommand \@href[1]{\@@startlink{#1}\@@href}%
\providecommand \@@href[1]{\endgroup#1\@@endlink}%
\providecommand \@sanitize@url [0]{\catcode `\\12\catcode `\$12\catcode
  `\&12\catcode `\#12\catcode `\^12\catcode `\_12\catcode `\%12\relax}%
\providecommand \@@startlink[1]{}%
\providecommand \@@endlink[0]{}%
\providecommand \url  [0]{\begingroup\@sanitize@url \@url }%
\providecommand \@url [1]{\endgroup\@href {#1}{\urlprefix }}%
\providecommand \urlprefix  [0]{URL }%
\providecommand \Eprint [0]{\href }%
\providecommand \doibase [0]{https://doi.org/}%
\providecommand \selectlanguage [0]{\@gobble}%
\providecommand \bibinfo  [0]{\@secondoftwo}%
\providecommand \bibfield  [0]{\@secondoftwo}%
\providecommand \translation [1]{[#1]}%
\providecommand \BibitemOpen [0]{}%
\providecommand \bibitemStop [0]{}%
\providecommand \bibitemNoStop [0]{.\EOS\space}%
\providecommand \EOS [0]{\spacefactor3000\relax}%
\providecommand \BibitemShut  [1]{\csname bibitem#1\endcsname}%
\let\auto@bib@innerbib\@empty
\bibitem [{\citenamefont
  {Sanmart{\'\i}n~Losada}(1984)}]{sanmartin1984botafumeiro}%
  \BibitemOpen
  \bibfield  {author} {\bibinfo {author} {\bibfnamefont {J.~R.}\ \bibnamefont
  {Sanmart{\'\i}n~Losada}},\ }\href@noop {} {\bibfield  {journal} {\bibinfo
  {journal} {American Journal of Physics}\ }\textbf {\bibinfo {volume} {52}},\
  \bibinfo {pages} {937} (\bibinfo {year} {1984})}\BibitemShut {NoStop}%
\bibitem [{\citenamefont {Grandi}\ \emph {et~al.}(2021)\citenamefont {Grandi},
  \citenamefont {Proti{\`e}re},\ and\ \citenamefont
  {Lazarus}}]{grandi2021enhancing}%
  \BibitemOpen
  \bibfield  {author} {\bibinfo {author} {\bibfnamefont {A.~A.}\ \bibnamefont
  {Grandi}}, \bibinfo {author} {\bibfnamefont {S.}~\bibnamefont
  {Proti{\`e}re}},\ and\ \bibinfo {author} {\bibfnamefont {A.}~\bibnamefont
  {Lazarus}},\ }\href@noop {} {\bibfield  {journal} {\bibinfo  {journal}
  {Extreme Mechanics Letters}\ ,\ \bibinfo {pages} {101195}} (\bibinfo {year}
  {2021})}\BibitemShut {NoStop}%
\bibitem [{\citenamefont {Faraday}(1831)}]{faraday1831xvii}%
  \BibitemOpen
  \bibfield  {author} {\bibinfo {author} {\bibfnamefont {M.}~\bibnamefont
  {Faraday}},\ }\href@noop {} {\bibfield  {journal} {\bibinfo  {journal}
  {Philosophical transactions of the Royal Society of London}\ ,\ \bibinfo
  {pages} {299}} (\bibinfo {year} {1831})}\BibitemShut {NoStop}%
\bibitem [{\citenamefont {Benjamin}\ and\ \citenamefont
  {Ursell}(1954)}]{benjamin1954stability}%
  \BibitemOpen
  \bibfield  {author} {\bibinfo {author} {\bibfnamefont {T.~B.}\ \bibnamefont
  {Benjamin}}\ and\ \bibinfo {author} {\bibfnamefont {F.~J.}\ \bibnamefont
  {Ursell}},\ }\href@noop {} {\bibfield  {journal} {\bibinfo  {journal}
  {Proceedings of the Royal Society of London. Series A. Mathematical and
  Physical Sciences}\ }\textbf {\bibinfo {volume} {225}},\ \bibinfo {pages}
  {505} (\bibinfo {year} {1954})}\BibitemShut {NoStop}%
\bibitem [{\citenamefont {Heugel}\ \emph {et~al.}(2019)\citenamefont {Heugel},
  \citenamefont {Oscity}, \citenamefont {Eichler}, \citenamefont {Zilberberg},\
  and\ \citenamefont {Chitra}}]{heugel2019classical}%
  \BibitemOpen
  \bibfield  {author} {\bibinfo {author} {\bibfnamefont {T.~L.}\ \bibnamefont
  {Heugel}}, \bibinfo {author} {\bibfnamefont {M.}~\bibnamefont {Oscity}},
  \bibinfo {author} {\bibfnamefont {A.}~\bibnamefont {Eichler}}, \bibinfo
  {author} {\bibfnamefont {O.}~\bibnamefont {Zilberberg}},\ and\ \bibinfo
  {author} {\bibfnamefont {R.}~\bibnamefont {Chitra}},\ }\href@noop {}
  {\bibfield  {journal} {\bibinfo  {journal} {Physical review letters}\
  }\textbf {\bibinfo {volume} {123}},\ \bibinfo {pages} {124301} (\bibinfo
  {year} {2019})}\BibitemShut {NoStop}%
\bibitem [{\citenamefont {Zhang}\ \emph {et~al.}(2017)\citenamefont {Zhang},
  \citenamefont {Hess}, \citenamefont {Kyprianidis}, \citenamefont {Becker},
  \citenamefont {Lee}, \citenamefont {Smith}, \citenamefont {Pagano},
  \citenamefont {Potirniche}, \citenamefont {Potter}, \citenamefont
  {Vishwanath} \emph {et~al.}}]{zhang2017observation}%
  \BibitemOpen
  \bibfield  {author} {\bibinfo {author} {\bibfnamefont {J.}~\bibnamefont
  {Zhang}}, \bibinfo {author} {\bibfnamefont {P.~W.}\ \bibnamefont {Hess}},
  \bibinfo {author} {\bibfnamefont {A.}~\bibnamefont {Kyprianidis}}, \bibinfo
  {author} {\bibfnamefont {P.}~\bibnamefont {Becker}}, \bibinfo {author}
  {\bibfnamefont {A.}~\bibnamefont {Lee}}, \bibinfo {author} {\bibfnamefont
  {J.}~\bibnamefont {Smith}}, \bibinfo {author} {\bibfnamefont
  {G.}~\bibnamefont {Pagano}}, \bibinfo {author} {\bibfnamefont {I.-D.}\
  \bibnamefont {Potirniche}}, \bibinfo {author} {\bibfnamefont {A.~C.}\
  \bibnamefont {Potter}}, \bibinfo {author} {\bibfnamefont {A.}~\bibnamefont
  {Vishwanath}}, \emph {et~al.},\ }\href@noop {} {\bibfield  {journal}
  {\bibinfo  {journal} {Nature}\ }\textbf {\bibinfo {volume} {543}},\ \bibinfo
  {pages} {217} (\bibinfo {year} {2017})}\BibitemShut {NoStop}%
\bibitem [{\citenamefont {Smith}\ and\ \citenamefont
  {Blackburn}(1992)}]{smith1992experimental}%
  \BibitemOpen
  \bibfield  {author} {\bibinfo {author} {\bibfnamefont {H.}~\bibnamefont
  {Smith}}\ and\ \bibinfo {author} {\bibfnamefont {J.~A.}\ \bibnamefont
  {Blackburn}},\ }\href@noop {} {\bibfield  {journal} {\bibinfo  {journal}
  {American Journal of Physics}\ }\textbf {\bibinfo {volume} {60}},\ \bibinfo
  {pages} {909} (\bibinfo {year} {1992})}\BibitemShut {NoStop}%
\bibitem [{\citenamefont {Acheson}(1993)}]{acheson1993upside}%
  \BibitemOpen
  \bibfield  {author} {\bibinfo {author} {\bibfnamefont {D.}~\bibnamefont
  {Acheson}},\ }\href@noop {} {\bibfield  {journal} {\bibinfo  {journal}
  {Nature}\ }\textbf {\bibinfo {volume} {366}},\ \bibinfo {pages} {215}
  (\bibinfo {year} {1993})}\BibitemShut {NoStop}%
\bibitem [{\citenamefont {Grandi}\ \emph {et~al.}(2023)\citenamefont {Grandi},
  \citenamefont {Proti{\`e}re},\ and\ \citenamefont {Lazarus}}]{grandi2023new}%
  \BibitemOpen
  \bibfield  {author} {\bibinfo {author} {\bibfnamefont {A.~A.}\ \bibnamefont
  {Grandi}}, \bibinfo {author} {\bibfnamefont {S.}~\bibnamefont
  {Proti{\`e}re}},\ and\ \bibinfo {author} {\bibfnamefont {A.}~\bibnamefont
  {Lazarus}},\ }\href@noop {} {\bibfield  {journal} {\bibinfo  {journal}
  {Nonlinear Dynamics}\ }\textbf {\bibinfo {volume} {111}},\ \bibinfo {pages}
  {12339} (\bibinfo {year} {2023})}\BibitemShut {NoStop}%
\bibitem [{\citenamefont {Audoly}\ and\ \citenamefont
  {Pomeau}(2010)}]{audoly2000elasticity}%
  \BibitemOpen
  \bibfield  {author} {\bibinfo {author} {\bibfnamefont {B.}~\bibnamefont
  {Audoly}}\ and\ \bibinfo {author} {\bibfnamefont {Y.}~\bibnamefont
  {Pomeau}},\ }\href@noop {} {\emph {\bibinfo {title} {Elasticity and geometry:
  From hair curls to the nonlinear response of shells}}}\ (\bibinfo
  {publisher} {Oxford press},\ \bibinfo {year} {2010})\BibitemShut {NoStop}%
\bibitem [{\citenamefont {Bazant}\ and\ \citenamefont
  {Cedolin}(2010)}]{bazant2010stability}%
  \BibitemOpen
  \bibfield  {author} {\bibinfo {author} {\bibfnamefont {Z.~P.}\ \bibnamefont
  {Bazant}}\ and\ \bibinfo {author} {\bibfnamefont {L.}~\bibnamefont
  {Cedolin}},\ }\href@noop {} {\emph {\bibinfo {title} {Stability of
  structures: elastic, inelastic, fracture and damage theories}}}\ (\bibinfo
  {publisher} {World Scientific},\ \bibinfo {year} {2010})\BibitemShut
  {NoStop}%
\bibitem [{\citenamefont {Krieger}(2012)}]{krieger2012extreme}%
  \BibitemOpen
  \bibfield  {author} {\bibinfo {author} {\bibfnamefont {K.}~\bibnamefont
  {Krieger}},\ }\href@noop {} {\bibfield  {journal} {\bibinfo  {journal}
  {Nature News}\ }\textbf {\bibinfo {volume} {488}},\ \bibinfo {pages} {146}
  (\bibinfo {year} {2012})}\BibitemShut {NoStop}%
\bibitem [{\citenamefont {Reis}(2015)}]{reis2015perspective}%
  \BibitemOpen
  \bibfield  {author} {\bibinfo {author} {\bibfnamefont {P.~M.}\ \bibnamefont
  {Reis}},\ }\href@noop {} {\bibfield  {journal} {\bibinfo  {journal} {Journal
  of Applied Mechanics}\ }\textbf {\bibinfo {volume} {82}},\ \bibinfo {pages}
  {111001} (\bibinfo {year} {2015})}\BibitemShut {NoStop}%
\bibitem [{\citenamefont {Shim}\ \emph {et~al.}(2012)\citenamefont {Shim},
  \citenamefont {Perdigou}, \citenamefont {Chen}, \citenamefont {Bertoldi},\
  and\ \citenamefont {Reis}}]{shim2012buckling}%
  \BibitemOpen
  \bibfield  {author} {\bibinfo {author} {\bibfnamefont {J.}~\bibnamefont
  {Shim}}, \bibinfo {author} {\bibfnamefont {C.}~\bibnamefont {Perdigou}},
  \bibinfo {author} {\bibfnamefont {E.~R.}\ \bibnamefont {Chen}}, \bibinfo
  {author} {\bibfnamefont {K.}~\bibnamefont {Bertoldi}},\ and\ \bibinfo
  {author} {\bibfnamefont {P.~M.}\ \bibnamefont {Reis}},\ }\href@noop {}
  {\bibfield  {journal} {\bibinfo  {journal} {Proceedings of the National
  Academy of Sciences}\ }\textbf {\bibinfo {volume} {109}},\ \bibinfo {pages}
  {5978} (\bibinfo {year} {2012})}\BibitemShut {NoStop}%
\bibitem [{\citenamefont {Lazarus}\ and\ \citenamefont
  {Reis}(2015)}]{lazarus2015soft}%
  \BibitemOpen
  \bibfield  {author} {\bibinfo {author} {\bibfnamefont {A.}~\bibnamefont
  {Lazarus}}\ and\ \bibinfo {author} {\bibfnamefont {P.~M.}\ \bibnamefont
  {Reis}},\ }\href@noop {} {\bibfield  {journal} {\bibinfo  {journal} {Advanced
  Engineering Materials}\ }\textbf {\bibinfo {volume} {17}},\ \bibinfo {pages}
  {815} (\bibinfo {year} {2015})}\BibitemShut {NoStop}%
\bibitem [{\citenamefont {Risso}\ \emph {et~al.}(2021)\citenamefont {Risso},
  \citenamefont {Sakovsky},\ and\ \citenamefont {Ermanni}}]{RISSO2021113946}%
  \BibitemOpen
  \bibfield  {author} {\bibinfo {author} {\bibfnamefont {G.}~\bibnamefont
  {Risso}}, \bibinfo {author} {\bibfnamefont {M.}~\bibnamefont {Sakovsky}},\
  and\ \bibinfo {author} {\bibfnamefont {P.}~\bibnamefont {Ermanni}},\ }\href
  {https://doi.org/https://doi.org/10.1016/j.compstruct.2021.113946} {\bibfield
   {journal} {\bibinfo  {journal} {Composite Structures}\ }\textbf {\bibinfo
  {volume} {268}},\ \bibinfo {pages} {113946} (\bibinfo {year}
  {2021})}\BibitemShut {NoStop}%
\bibitem [{\citenamefont {Yang}\ \emph {et~al.}(2024)\citenamefont {Yang},
  \citenamefont {Read}, \citenamefont {Sbai}, \citenamefont {Zareei},
  \citenamefont {Forte}, \citenamefont {Melancon},\ and\ \citenamefont
  {Bertoldi}}]{yang2024complex}%
  \BibitemOpen
  \bibfield  {author} {\bibinfo {author} {\bibfnamefont {Y.}~\bibnamefont
  {Yang}}, \bibinfo {author} {\bibfnamefont {H.}~\bibnamefont {Read}}, \bibinfo
  {author} {\bibfnamefont {M.}~\bibnamefont {Sbai}}, \bibinfo {author}
  {\bibfnamefont {A.}~\bibnamefont {Zareei}}, \bibinfo {author} {\bibfnamefont
  {A.~E.}\ \bibnamefont {Forte}}, \bibinfo {author} {\bibfnamefont
  {D.}~\bibnamefont {Melancon}},\ and\ \bibinfo {author} {\bibfnamefont
  {K.}~\bibnamefont {Bertoldi}},\ }\href@noop {} {\bibfield  {journal}
  {\bibinfo  {journal} {Advanced Materials}\ }\textbf {\bibinfo {volume}
  {36}},\ \bibinfo {pages} {2406611} (\bibinfo {year} {2024})}\BibitemShut
  {NoStop}%
\bibitem [{\citenamefont {Lee}\ \emph {et~al.}(2014)\citenamefont {Lee},
  \citenamefont {Zhang}, \citenamefont {Cho}, \citenamefont {Cui},
  \citenamefont {Van~der Spiegel}, \citenamefont {Engheta},\ and\ \citenamefont
  {Yang}}]{lee2014tilted}%
  \BibitemOpen
  \bibfield  {author} {\bibinfo {author} {\bibfnamefont {E.}~\bibnamefont
  {Lee}}, \bibinfo {author} {\bibfnamefont {M.}~\bibnamefont {Zhang}}, \bibinfo
  {author} {\bibfnamefont {Y.}~\bibnamefont {Cho}}, \bibinfo {author}
  {\bibfnamefont {Y.}~\bibnamefont {Cui}}, \bibinfo {author} {\bibfnamefont
  {J.}~\bibnamefont {Van~der Spiegel}}, \bibinfo {author} {\bibfnamefont
  {N.}~\bibnamefont {Engheta}},\ and\ \bibinfo {author} {\bibfnamefont
  {S.}~\bibnamefont {Yang}},\ }\href@noop {} {\bibfield  {journal} {\bibinfo
  {journal} {Advanced Materials}\ }\textbf {\bibinfo {volume} {26}},\ \bibinfo
  {pages} {4127} (\bibinfo {year} {2014})}\BibitemShut {NoStop}%
\bibitem [{\citenamefont {Doar{\'e}}\ and\ \citenamefont
  {Michelin}(2011)}]{doare2011piezoelectric}%
  \BibitemOpen
  \bibfield  {author} {\bibinfo {author} {\bibfnamefont {O.}~\bibnamefont
  {Doar{\'e}}}\ and\ \bibinfo {author} {\bibfnamefont {S.}~\bibnamefont
  {Michelin}},\ }\href@noop {} {\bibfield  {journal} {\bibinfo  {journal}
  {Journal of Fluids and Structures}\ }\textbf {\bibinfo {volume} {27}},\
  \bibinfo {pages} {1357} (\bibinfo {year} {2011})}\BibitemShut {NoStop}%
\bibitem [{\citenamefont {Michelin}\ and\ \citenamefont
  {Doar{\'e}}(2013)}]{michelin2013energy}%
  \BibitemOpen
  \bibfield  {author} {\bibinfo {author} {\bibfnamefont {S.}~\bibnamefont
  {Michelin}}\ and\ \bibinfo {author} {\bibfnamefont {O.}~\bibnamefont
  {Doar{\'e}}},\ }\href@noop {} {\bibfield  {journal} {\bibinfo  {journal}
  {Journal of Fluid Mechanics}\ }\textbf {\bibinfo {volume} {714}},\ \bibinfo
  {pages} {489} (\bibinfo {year} {2013})}\BibitemShut {NoStop}%
\bibitem [{\citenamefont {Kang}\ \emph {et~al.}(2013)\citenamefont {Kang},
  \citenamefont {Shan}, \citenamefont {Noorduin}, \citenamefont {Khan},
  \citenamefont {Aizenberg},\ and\ \citenamefont
  {Bertoldi}}]{kang2013buckling}%
  \BibitemOpen
  \bibfield  {author} {\bibinfo {author} {\bibfnamefont {S.~H.}\ \bibnamefont
  {Kang}}, \bibinfo {author} {\bibfnamefont {S.}~\bibnamefont {Shan}}, \bibinfo
  {author} {\bibfnamefont {W.~L.}\ \bibnamefont {Noorduin}}, \bibinfo {author}
  {\bibfnamefont {M.}~\bibnamefont {Khan}}, \bibinfo {author} {\bibfnamefont
  {J.}~\bibnamefont {Aizenberg}},\ and\ \bibinfo {author} {\bibfnamefont
  {K.}~\bibnamefont {Bertoldi}},\ }\href@noop {} {\bibfield  {journal}
  {\bibinfo  {journal} {Advanced materials}\ } (\bibinfo {year}
  {2013})}\BibitemShut {NoStop}%
\bibitem [{\citenamefont {Deng}\ \emph {et~al.}(2020)\citenamefont {Deng},
  \citenamefont {Yu}, \citenamefont {Forte}, \citenamefont {Tournat},\ and\
  \citenamefont {Bertoldi}}]{deng2020characterization}%
  \BibitemOpen
  \bibfield  {author} {\bibinfo {author} {\bibfnamefont {B.}~\bibnamefont
  {Deng}}, \bibinfo {author} {\bibfnamefont {S.}~\bibnamefont {Yu}}, \bibinfo
  {author} {\bibfnamefont {A.~E.}\ \bibnamefont {Forte}}, \bibinfo {author}
  {\bibfnamefont {V.}~\bibnamefont {Tournat}},\ and\ \bibinfo {author}
  {\bibfnamefont {K.}~\bibnamefont {Bertoldi}},\ }\href@noop {} {\bibfield
  {journal} {\bibinfo  {journal} {Proceedings of the National Academy of
  Sciences}\ }\textbf {\bibinfo {volume} {117}},\ \bibinfo {pages} {31002}
  (\bibinfo {year} {2020})}\BibitemShut {NoStop}%
\bibitem [{\citenamefont {Jin}\ \emph {et~al.}(2020)\citenamefont {Jin},
  \citenamefont {Khajehtourian}, \citenamefont {Mueller}, \citenamefont
  {Rafsanjani}, \citenamefont {Tournat}, \citenamefont {Bertoldi},\ and\
  \citenamefont {Kochmann}}]{jin2020guided}%
  \BibitemOpen
  \bibfield  {author} {\bibinfo {author} {\bibfnamefont {L.}~\bibnamefont
  {Jin}}, \bibinfo {author} {\bibfnamefont {R.}~\bibnamefont {Khajehtourian}},
  \bibinfo {author} {\bibfnamefont {J.}~\bibnamefont {Mueller}}, \bibinfo
  {author} {\bibfnamefont {A.}~\bibnamefont {Rafsanjani}}, \bibinfo {author}
  {\bibfnamefont {V.}~\bibnamefont {Tournat}}, \bibinfo {author} {\bibfnamefont
  {K.}~\bibnamefont {Bertoldi}},\ and\ \bibinfo {author} {\bibfnamefont
  {D.~M.}\ \bibnamefont {Kochmann}},\ }\href@noop {} {\bibfield  {journal}
  {\bibinfo  {journal} {Proceedings of the National Academy of Sciences}\
  }\textbf {\bibinfo {volume} {117}},\ \bibinfo {pages} {2319} (\bibinfo {year}
  {2020})}\BibitemShut {NoStop}%
\bibitem [{\citenamefont {Nadkarni}\ \emph {et~al.}(2016)\citenamefont
  {Nadkarni}, \citenamefont {Arrieta}, \citenamefont {Chong}, \citenamefont
  {Kochmann},\ and\ \citenamefont {Daraio}}]{nadkarni2016unidirectional}%
  \BibitemOpen
  \bibfield  {author} {\bibinfo {author} {\bibfnamefont {N.}~\bibnamefont
  {Nadkarni}}, \bibinfo {author} {\bibfnamefont {A.~F.}\ \bibnamefont
  {Arrieta}}, \bibinfo {author} {\bibfnamefont {C.}~\bibnamefont {Chong}},
  \bibinfo {author} {\bibfnamefont {D.~M.}\ \bibnamefont {Kochmann}},\ and\
  \bibinfo {author} {\bibfnamefont {C.}~\bibnamefont {Daraio}},\ }\href@noop {}
  {\bibfield  {journal} {\bibinfo  {journal} {Physical review letters}\
  }\textbf {\bibinfo {volume} {116}},\ \bibinfo {pages} {244501} (\bibinfo
  {year} {2016})}\BibitemShut {NoStop}%
\bibitem [{\citenamefont {Brau}\ \emph {et~al.}(2011)\citenamefont {Brau},
  \citenamefont {Vandeparre}, \citenamefont {Sabbah}, \citenamefont {Poulard},
  \citenamefont {Boudaoud},\ and\ \citenamefont {Damman}}]{brau2011multiple}%
  \BibitemOpen
  \bibfield  {author} {\bibinfo {author} {\bibfnamefont {F.}~\bibnamefont
  {Brau}}, \bibinfo {author} {\bibfnamefont {H.}~\bibnamefont {Vandeparre}},
  \bibinfo {author} {\bibfnamefont {A.}~\bibnamefont {Sabbah}}, \bibinfo
  {author} {\bibfnamefont {C.}~\bibnamefont {Poulard}}, \bibinfo {author}
  {\bibfnamefont {A.}~\bibnamefont {Boudaoud}},\ and\ \bibinfo {author}
  {\bibfnamefont {P.}~\bibnamefont {Damman}},\ }\href@noop {} {\bibfield
  {journal} {\bibinfo  {journal} {Nature Physics}\ }\textbf {\bibinfo {volume}
  {7}},\ \bibinfo {pages} {56} (\bibinfo {year} {2011})}\BibitemShut {NoStop}%
\bibitem [{\citenamefont {Brau}\ \emph {et~al.}(2013)\citenamefont {Brau},
  \citenamefont {Damman}, \citenamefont {Diamant},\ and\ \citenamefont
  {Witten}}]{brau2013wrinkle}%
  \BibitemOpen
  \bibfield  {author} {\bibinfo {author} {\bibfnamefont {F.}~\bibnamefont
  {Brau}}, \bibinfo {author} {\bibfnamefont {P.}~\bibnamefont {Damman}},
  \bibinfo {author} {\bibfnamefont {H.}~\bibnamefont {Diamant}},\ and\ \bibinfo
  {author} {\bibfnamefont {T.~A.}\ \bibnamefont {Witten}},\ }\href@noop {}
  {\bibfield  {journal} {\bibinfo  {journal} {Soft Matter}\ }\textbf {\bibinfo
  {volume} {9}},\ \bibinfo {pages} {8177} (\bibinfo {year} {2013})}\BibitemShut
  {NoStop}%
\bibitem [{\citenamefont {Lagrange}\ and\ \citenamefont
  {Averbuch}(2012)}]{lagrange2012solution}%
  \BibitemOpen
  \bibfield  {author} {\bibinfo {author} {\bibfnamefont {R.}~\bibnamefont
  {Lagrange}}\ and\ \bibinfo {author} {\bibfnamefont {D.}~\bibnamefont
  {Averbuch}},\ }\href@noop {} {\bibfield  {journal} {\bibinfo  {journal}
  {International Journal of Mechanical Sciences}\ }\textbf {\bibinfo {volume}
  {63}},\ \bibinfo {pages} {48} (\bibinfo {year} {2012})}\BibitemShut {NoStop}%
\bibitem [{\citenamefont {Richards}(2012)}]{richards2012analysis}%
  \BibitemOpen
  \bibfield  {author} {\bibinfo {author} {\bibfnamefont {J.~A.}\ \bibnamefont
  {Richards}},\ }\href@noop {} {\emph {\bibinfo {title} {Analysis of
  periodically time-varying systems}}}\ (\bibinfo  {publisher} {Springer
  Science \& Business Media},\ \bibinfo {year} {2012})\BibitemShut {NoStop}%
\bibitem [{\citenamefont {Hill}(1886)}]{hill1886part}%
  \BibitemOpen
  \bibfield  {author} {\bibinfo {author} {\bibfnamefont {G.~W.}\ \bibnamefont
  {Hill}},\ }\href@noop {} {\bibfield  {journal} {\bibinfo  {journal} {Acta
  Mathematica}\ }\textbf {\bibinfo {volume} {8}},\ \bibinfo {pages} {3}
  (\bibinfo {year} {1886})}\BibitemShut {NoStop}%
\bibitem [{\citenamefont {Deconinck}\ \emph {et~al.}(2007)\citenamefont
  {Deconinck}, \citenamefont {Kiyak}, \citenamefont {Carter},\ and\
  \citenamefont {Kutz}}]{deconinck2007spectruw}%
  \BibitemOpen
  \bibfield  {author} {\bibinfo {author} {\bibfnamefont {B.}~\bibnamefont
  {Deconinck}}, \bibinfo {author} {\bibfnamefont {F.}~\bibnamefont {Kiyak}},
  \bibinfo {author} {\bibfnamefont {J.~D.}\ \bibnamefont {Carter}},\ and\
  \bibinfo {author} {\bibfnamefont {J.~N.}\ \bibnamefont {Kutz}},\ }\href@noop
  {} {\bibfield  {journal} {\bibinfo  {journal} {Mathematics and Computers in
  Simulation}\ }\textbf {\bibinfo {volume} {74}},\ \bibinfo {pages} {370}
  (\bibinfo {year} {2007})}\BibitemShut {NoStop}%
\bibitem [{\citenamefont {Bentvelsen}\ and\ \citenamefont
  {Lazarus}(2018)}]{bentvelsen2017modal}%
  \BibitemOpen
  \bibfield  {author} {\bibinfo {author} {\bibfnamefont {B.}~\bibnamefont
  {Bentvelsen}}\ and\ \bibinfo {author} {\bibfnamefont {A.}~\bibnamefont
  {Lazarus}},\ }\href@noop {} {\bibfield  {journal} {\bibinfo  {journal}
  {Nonlinear Dynamics}\ }\textbf {\bibinfo {volume} {91}},\ \bibinfo {pages}
  {1349} (\bibinfo {year} {2018})}\BibitemShut {NoStop}%
\bibitem [{\citenamefont {Mora}\ and\ \citenamefont
  {Boudaoud}(2006)}]{mora2006buckling}%
  \BibitemOpen
  \bibfield  {author} {\bibinfo {author} {\bibfnamefont {T.}~\bibnamefont
  {Mora}}\ and\ \bibinfo {author} {\bibfnamefont {A.}~\bibnamefont
  {Boudaoud}},\ }\href@noop {} {\bibfield  {journal} {\bibinfo  {journal} {The
  European Physical Journal E}\ }\textbf {\bibinfo {volume} {20}},\ \bibinfo
  {pages} {119} (\bibinfo {year} {2006})}\BibitemShut {NoStop}%
\bibitem [{\citenamefont {Si{\'e}fert}\ \emph {et~al.}(2021)\citenamefont
  {Si{\'e}fert}, \citenamefont {Cattaud}, \citenamefont {Reyssat},
  \citenamefont {Roman},\ and\ \citenamefont {Bico}}]{Siefert2021stretch}%
  \BibitemOpen
  \bibfield  {author} {\bibinfo {author} {\bibfnamefont {E.}~\bibnamefont
  {Si{\'e}fert}}, \bibinfo {author} {\bibfnamefont {N.}~\bibnamefont
  {Cattaud}}, \bibinfo {author} {\bibfnamefont {E.}~\bibnamefont {Reyssat}},
  \bibinfo {author} {\bibfnamefont {B.}~\bibnamefont {Roman}},\ and\ \bibinfo
  {author} {\bibfnamefont {J.}~\bibnamefont {Bico}},\ }\href@noop {} {\bibfield
   {journal} {\bibinfo  {journal} {Physical review letters}\ }\textbf {\bibinfo
  {volume} {127}},\ \bibinfo {pages} {168002} (\bibinfo {year}
  {2021})}\BibitemShut {NoStop}%
\bibitem [{\citenamefont {Brillouin}(1946)}]{brillouin1946wave}%
  \BibitemOpen
  \bibfield  {author} {\bibinfo {author} {\bibfnamefont {L.}~\bibnamefont
  {Brillouin}},\ }\href@noop {} {\emph {\bibinfo {title} {Wave propagation in
  periodic structures}}},\ \bibinfo {edition} {1st}\ ed.\ (\bibinfo
  {publisher} {McGraw-Hill New York},\ \bibinfo {year} {1946})\BibitemShut
  {NoStop}%
\bibitem [{\citenamefont {Chen}\ and\ \citenamefont
  {Hutchinson}(2004)}]{chen2004herringbone}%
  \BibitemOpen
  \bibfield  {author} {\bibinfo {author} {\bibfnamefont {X.}~\bibnamefont
  {Chen}}\ and\ \bibinfo {author} {\bibfnamefont {J.~W.}\ \bibnamefont
  {Hutchinson}},\ }\href@noop {} {\bibfield  {journal} {\bibinfo  {journal}
  {Journal of Applied Mechanics}\ }\textbf {\bibinfo {volume} {71}},\ \bibinfo
  {pages} {597} (\bibinfo {year} {2004})}\BibitemShut {NoStop}%
\bibitem [{\citenamefont {Lee}\ \emph {et~al.}(2017)\citenamefont {Lee},
  \citenamefont {Bae}, \citenamefont {Lee},\ and\ \citenamefont
  {Yoon}}]{lee2017mechano}%
  \BibitemOpen
  \bibfield  {author} {\bibinfo {author} {\bibfnamefont {H.}~\bibnamefont
  {Lee}}, \bibinfo {author} {\bibfnamefont {J.~G.}\ \bibnamefont {Bae}},
  \bibinfo {author} {\bibfnamefont {W.~B.}\ \bibnamefont {Lee}},\ and\ \bibinfo
  {author} {\bibfnamefont {H.}~\bibnamefont {Yoon}},\ }\href@noop {} {\bibfield
   {journal} {\bibinfo  {journal} {Soft matter}\ }\textbf {\bibinfo {volume}
  {13}},\ \bibinfo {pages} {8357} (\bibinfo {year} {2017})}\BibitemShut
  {NoStop}%
\bibitem [{\citenamefont {Wu}\ \emph {et~al.}(2022)\citenamefont {Wu},
  \citenamefont {Huang}, \citenamefont {Zhang}, \citenamefont {Wang},
  \citenamefont {Pei}, \citenamefont {Zhao},\ and\ \citenamefont
  {Fang}}]{wu2022initiation}%
  \BibitemOpen
  \bibfield  {author} {\bibinfo {author} {\bibfnamefont {D.}~\bibnamefont
  {Wu}}, \bibinfo {author} {\bibfnamefont {Y.}~\bibnamefont {Huang}}, \bibinfo
  {author} {\bibfnamefont {Q.}~\bibnamefont {Zhang}}, \bibinfo {author}
  {\bibfnamefont {P.}~\bibnamefont {Wang}}, \bibinfo {author} {\bibfnamefont
  {Y.}~\bibnamefont {Pei}}, \bibinfo {author} {\bibfnamefont {Z.}~\bibnamefont
  {Zhao}},\ and\ \bibinfo {author} {\bibfnamefont {D.}~\bibnamefont {Fang}},\
  }\href@noop {} {\bibfield  {journal} {\bibinfo  {journal} {Journal of the
  Mechanics and Physics of Solids}\ }\textbf {\bibinfo {volume} {162}},\
  \bibinfo {pages} {104838} (\bibinfo {year} {2022})}\BibitemShut {NoStop}%
\bibitem [{\citenamefont {L{\"o}we}\ \emph {et~al.}(2005)\citenamefont
  {L{\"o}we}, \citenamefont {Zhang},\ and\ \citenamefont
  {Kovacs}}]{lowe2005dielectric}%
  \BibitemOpen
  \bibfield  {author} {\bibinfo {author} {\bibfnamefont {C.}~\bibnamefont
  {L{\"o}we}}, \bibinfo {author} {\bibfnamefont {X.}~\bibnamefont {Zhang}},\
  and\ \bibinfo {author} {\bibfnamefont {G.}~\bibnamefont {Kovacs}},\
  }\href@noop {} {\bibfield  {journal} {\bibinfo  {journal} {Advanced
  engineering materials}\ }\textbf {\bibinfo {volume} {7}},\ \bibinfo {pages}
  {361} (\bibinfo {year} {2005})}\BibitemShut {NoStop}%
\bibitem [{\citenamefont {Florijn}\ \emph {et~al.}(2014)\citenamefont
  {Florijn}, \citenamefont {Coulais},\ and\ \citenamefont {van
  Hecke}}]{florijn2014programmable}%
  \BibitemOpen
  \bibfield  {author} {\bibinfo {author} {\bibfnamefont {B.}~\bibnamefont
  {Florijn}}, \bibinfo {author} {\bibfnamefont {C.}~\bibnamefont {Coulais}},\
  and\ \bibinfo {author} {\bibfnamefont {M.}~\bibnamefont {van Hecke}},\
  }\href@noop {} {\emph {\bibinfo {title} {Programmable mechanical
  metamaterials}}},\ Vol.\ \bibinfo {volume} {113}\ (\bibinfo  {publisher}
  {APS},\ \bibinfo {year} {2014})\BibitemShut {NoStop}%
\bibitem [{\citenamefont {Ghatak}\ \emph {et~al.}(2020)\citenamefont {Ghatak},
  \citenamefont {Brandenbourger}, \citenamefont {Van~Wezel},\ and\
  \citenamefont {Coulais}}]{ghatak2020observation}%
  \BibitemOpen
  \bibfield  {author} {\bibinfo {author} {\bibfnamefont {A.}~\bibnamefont
  {Ghatak}}, \bibinfo {author} {\bibfnamefont {M.}~\bibnamefont
  {Brandenbourger}}, \bibinfo {author} {\bibfnamefont {J.}~\bibnamefont
  {Van~Wezel}},\ and\ \bibinfo {author} {\bibfnamefont {C.}~\bibnamefont
  {Coulais}},\ }\href@noop {} {\bibfield  {journal} {\bibinfo  {journal}
  {Proceedings of the National Academy of Sciences}\ }\textbf {\bibinfo
  {volume} {117}},\ \bibinfo {pages} {29561} (\bibinfo {year}
  {2020})}\BibitemShut {NoStop}%
\bibitem [{\citenamefont {Lazarus}\ and\ \citenamefont
  {Tr{\'e}lat}(2025{\natexlab{a}})}]{lazarus2025optimal}%
  \BibitemOpen
  \bibfield  {author} {\bibinfo {author} {\bibfnamefont {A.}~\bibnamefont
  {Lazarus}}\ and\ \bibinfo {author} {\bibfnamefont {E.}~\bibnamefont
  {Tr{\'e}lat}},\ }\href@noop {} {\bibfield  {journal} {\bibinfo  {journal}
  {arXiv preprint arXiv:2508.00006}\ } (\bibinfo {year}
  {2025}{\natexlab{a}})}\BibitemShut {NoStop}%
\bibitem [{\citenamefont {Lazarus}\ and\ \citenamefont
  {Tr{\'e}lat}(2025{\natexlab{b}})}]{lazarus2025meaningful}%
  \BibitemOpen
  \bibfield  {author} {\bibinfo {author} {\bibfnamefont {A.}~\bibnamefont
  {Lazarus}}\ and\ \bibinfo {author} {\bibfnamefont {E.}~\bibnamefont
  {Tr{\'e}lat}},\ }\href@noop {} {\bibfield  {journal} {\bibinfo  {journal}
  {arXiv preprint arXiv:2507.13125}\ } (\bibinfo {year}
  {2025}{\natexlab{b}})}\BibitemShut {NoStop}%
\bibitem [{\citenamefont {Lazarus}\ \emph {et~al.}(2015)\citenamefont
  {Lazarus}, \citenamefont {Maurini},\ and\ \citenamefont
  {Neukirch}}]{lazarus2015stability}%
  \BibitemOpen
  \bibfield  {author} {\bibinfo {author} {\bibfnamefont {A.}~\bibnamefont
  {Lazarus}}, \bibinfo {author} {\bibfnamefont {C.}~\bibnamefont {Maurini}},\
  and\ \bibinfo {author} {\bibfnamefont {S.}~\bibnamefont {Neukirch}},\ }in\
  \href@noop {} {\emph {\bibinfo {booktitle} {Extremely Deformable
  Structures}}}\ (\bibinfo  {publisher} {Springer},\ \bibinfo {year} {2015})\
  Chap.~\bibinfo {chapter} {1}, pp.\ \bibinfo {pages} {1--53}\BibitemShut
  {NoStop}%
\bibitem [{\citenamefont {Calico}\ and\ \citenamefont
  {Wiesel}(1984)}]{calico1984control}%
  \BibitemOpen
  \bibfield  {author} {\bibinfo {author} {\bibfnamefont {R.~A.}\ \bibnamefont
  {Calico}}\ and\ \bibinfo {author} {\bibfnamefont {W.~E.}\ \bibnamefont
  {Wiesel}},\ }\href@noop {} {\bibfield  {journal} {\bibinfo  {journal}
  {Journal of Guidance, Control, and Dynamics}\ }\textbf {\bibinfo {volume}
  {7}},\ \bibinfo {pages} {671} (\bibinfo {year} {1984})}\BibitemShut {NoStop}%
\bibitem [{\citenamefont {Zhou}\ \emph {et~al.}(2004)\citenamefont {Zhou},
  \citenamefont {Hagiwara},\ and\ \citenamefont {Araki}}]{zhou2004spectral}%
  \BibitemOpen
  \bibfield  {author} {\bibinfo {author} {\bibfnamefont {J.}~\bibnamefont
  {Zhou}}, \bibinfo {author} {\bibfnamefont {T.}~\bibnamefont {Hagiwara}},\
  and\ \bibinfo {author} {\bibfnamefont {M.}~\bibnamefont {Araki}},\
  }\href@noop {} {\bibfield  {journal} {\bibinfo  {journal} {Systems \& control
  letters}\ }\textbf {\bibinfo {volume} {53}},\ \bibinfo {pages} {141}
  (\bibinfo {year} {2004})}\BibitemShut {NoStop}%
\bibitem [{\citenamefont {Liu}\ and\ \citenamefont
  {Bertoldi}(2015)}]{liu2015bloch}%
  \BibitemOpen
  \bibfield  {author} {\bibinfo {author} {\bibfnamefont {J.}~\bibnamefont
  {Liu}}\ and\ \bibinfo {author} {\bibfnamefont {K.}~\bibnamefont {Bertoldi}},\
  }\href@noop {} {\bibfield  {journal} {\bibinfo  {journal} {Proc. R. Soc. A}\
  }\textbf {\bibinfo {volume} {471}},\ \bibinfo {pages} {20150493} (\bibinfo
  {year} {2015})}\BibitemShut {NoStop}%
\bibitem [{\citenamefont {{\AA}berg}\ and\ \citenamefont
  {Gudmundson}(1997)}]{aaberg1997usage}%
  \BibitemOpen
  \bibfield  {author} {\bibinfo {author} {\bibfnamefont {M.}~\bibnamefont
  {{\AA}berg}}\ and\ \bibinfo {author} {\bibfnamefont {P.}~\bibnamefont
  {Gudmundson}},\ }\href@noop {} {\bibfield  {journal} {\bibinfo  {journal}
  {The Journal of the Acoustical Society of America}\ }\textbf {\bibinfo
  {volume} {102}},\ \bibinfo {pages} {2007} (\bibinfo {year}
  {1997})}\BibitemShut {NoStop}%
\end{thebibliography}%

\clearpage
\newpage
\onecolumngrid

\appendix

\section{Parametric Pendulum} \label{sec:inverted_pendulum}

\begin{figure}[h!]
    \centering
    \includegraphics{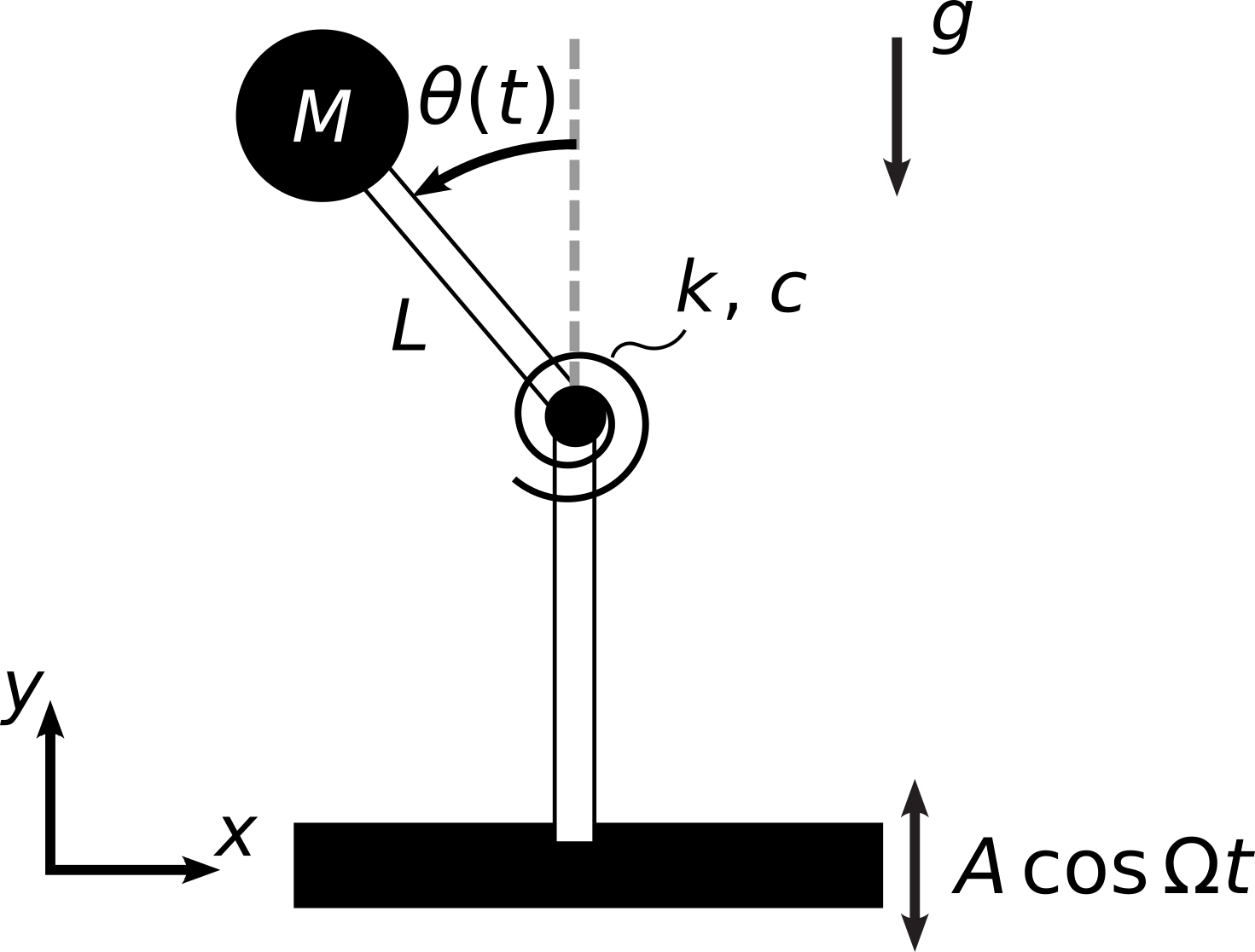}
    \caption{Schematic of an inverted pendulum with a shaking base. We study the stability and spectral response of $\theta(t)$ as a function of amplitude and period of the shaking.}
    \label{fig:si_inv_pen_sch}
\end{figure}

We begin by considering the inverted pendulum shown in Fig.~1 of the main text, or Fig.~\ref{fig:si_inv_pen_sch} here. 

This pendulum has mass $M$ and length $L$ and is connected to a rigid base via a rotational spring with stiffness $k$ and a rotational dashpot with damping coefficient $c$. Additionally, the base of this pendulum is oscillated by applying a harmonic displacement along the vertical direction with amplitude $A$ and frequency $\Omega$ 
\begin{equation}
    u_{\text{applied}}=A \cos{\Omega t}.
\end{equation}
The equation of motion for such a system is given by \cite{lazarus2015stability}. 


\begin{equation}\label{eqn:inv_pendem_nonlinear}
    \ddot{\theta} + \frac{c}{ML^2} \dot{\theta} + \frac{k}{ML^2} \theta + \left(\frac{A \Omega^2}{L} \cos{\Omega t} - \frac{g}{L} \right) \sin{\theta} = 0,
\end{equation}
where $g$ denotes the gravitational acceleration, $\theta$ is the angle from the vertical to the pendulum and $\dot{\theta}= \diff{\theta}{t}$. We see from Eq.~(\ref{eqn:inv_pendem_nonlinear}) that in the absence of forcing (i.e. $A = 0$) and for small oscillations this system's undamped natural frequency is

\begin{equation}
    \omega_0 = \left(\frac{k}{ML^2} - \frac{g}{L} \right)^{1/2}.
\end{equation}

Next, we nondimensionalize Eq.~(\ref{eqn:inv_pendem_nonlinear}) to reduce the number of independent variables from six ($c$, $k$, $M$, $L$, $A$, $\Omega$) to three. Toward this end, we begin by introducing the normalized time  
\begin{equation}\label{tau}
    \tau = \omega_0 t.
\end{equation}

Substitution of Eq.~(\ref{tau}) into Eq.~(\ref{eqn:inv_pendem_nonlinear}) yields

\begin{equation}\label{eqn:inv_pendulum_nondim_1}
    \omega_0^2 \theta'' + \frac{c}{ML^2} \omega_0 \theta' + \frac{k}{ML^2} \theta + \left(\frac{A \Omega^2}{L} \cos{\frac{\Omega}{\omega_0} \tau} - \frac{g}{L} \right) \sin{\theta} = 0,
\end{equation}

where $\theta' = \diff{\theta}{\tau}$ denotes a derivative with respect to $\tau$ rather than $t$. 
Next, we multiply both sides of Eq. (\ref{eqn:inv_pendulum_nondim_1}) by $1/\omega_0^2$ and linearize it about $\theta_0=0$ by assuming $\sin{\theta} \approx \theta$, yielding

\begin{align}
    \frac{1}{\omega_0^2} \left[ \omega_0^2 \theta'' + \frac{c}{ML^2} \omega_0 \theta' + \frac{k}{ML^2} \theta + \left(\frac{A \Omega^2}{L} \cos{\frac{\Omega}{\omega_0} \tau} - \frac{g}{L} \right) \theta \right] &= \frac{1}{\omega_0^2} \cdot 0 \label{test1}\\
    \theta'' + \frac{c}{ML^2} \frac{1}{\omega_0} \theta' + \frac{1}{\omega_0^2} \underbrace{\left(\frac{k}{ML^2} - \frac{g}{L} \right)}_{\omega_0^2} \theta + \frac{1}{\omega_0^2} \frac{A\Omega^2}{L} \cos{\frac{\Omega}{\omega_0} \tau} \, \theta &= 0  \label{test2}\\
    \theta'' + \frac{c}{\omega_0 ML^2} \theta' + \left(1 + \frac{A \Omega^2}{L \omega_0^2} \cos{\frac{\Omega}{\omega_0} \tau} \right) \theta &= 0. \label{eqn:inv_pendulum_nondim_2}
\end{align}

Eq. (\ref{eqn:inv_pendulum_nondim_2}) shows that the system's behavior is governed by three dimensionless quantities: the normalized driving frequency, 

\begin{equation}\label{eqn:omega_bar}
    \bar{\Omega}=\frac{\Omega}{ \omega_0 },
\end{equation}

the normalized amplitude, 

\begin{equation}\label{eqn:A_bar}
    \bar{A}=\frac{A\Omega^2}{L \omega_0^2},
\end{equation}

and the normalized damping factor,

\begin{equation}\label{eqn:C_bar}
    \bar{C} = \frac{c}{\omega_0 M L^2}.
\end{equation}

Substitution of Eqs.~(\ref{eqn:omega_bar}), (\ref{eqn:A_bar}) and (\ref{eqn:C_bar}) into Eq. (\ref{eqn:inv_pendulum_nondim_2}) yields 

\begin{equation} \label{eqn:inv_pend_nondim_fin}
    \theta'' + \bar{C} \theta' + \left(1 + \bar{A} \cos{\bar{\Omega} \tau} \right) \theta = 0,
\end{equation}
which is the linearized and non-dimensionalized equation of motion for the system about $\theta_0 = 0$.

Finally, we rewrite Eq.~(\ref{eqn:inv_pend_nondim_fin}) as a first order system of equations, which is the form used for Floquet analysis (discussed further in Section~\ref{sec:floquet_theory})

\begin{equation} \label{eqn:inv_pend_nondim_system}
    \begin{pmatrix}
        \theta'\\ \theta''
    \end{pmatrix} =
    \begin{pmatrix}
        0 & 1\\ -1 - \bar{A}\cos{\bar{\Omega} \tau} & -\bar{C}
    \end{pmatrix}
    \begin{pmatrix}
        \theta\\ \theta'
    \end{pmatrix}.
\end{equation}

\section{Modulated Winkler Foundation} \label{sec:winkler}

\begin{figure}[h!]
    \centering
    \includegraphics{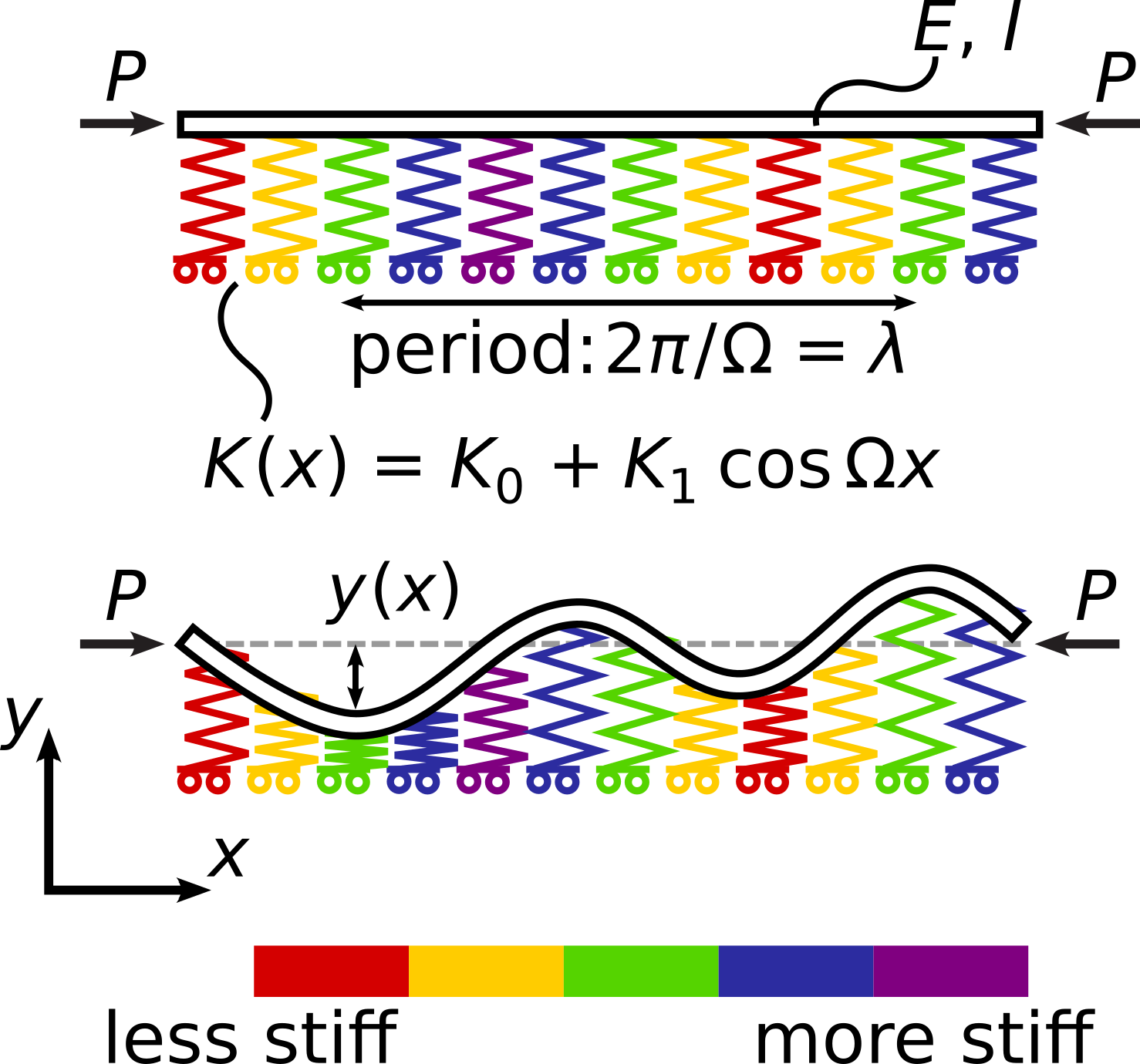}
    \caption{Schematic of the modulated Winkler foundation. We study the buckling patters $y(x)$ as a function of period and modulation of foundation stiffness.}
    \label{fig:si_winkler_sch}
\end{figure}

Here we consider an elastic beam with bending modulus $EI$ on a linear Winkler foundation with a periodic stiffness per unit length (see Fig.~1 of main text or Fig.~\ref{fig:si_winkler_sch} here.)
\begin{equation}
    K(x) = K_0 + K_1 \cos{\Omega x},
\end{equation}

where $K_0$ is the base stiffness, $K_1$ is the amplitude of modulation, and $\Omega$ is the frequency of modulation, with an associated wavelength $\lambda = 2\pi / \Omega$.

We investigate the buckling patterns of the beam---the $y$-deflection---under an axial load $P$.
When modeling the beam as an inextensible Euler–Bernoulli beam, the linearized equilibrium equation governing its transverse deflection about $y(x) = 0$ is given by \cite{brau2013wrinkle,lagrange2012solution}

\begin{equation}\label{eq:Eq1B}
EI\diff[4]{y}{x} + P \diff[2]{y}{x} + (K_0 + K_1 \cos{\Omega x}) \, y(x) = 0.
\end{equation}
Like the non-dimensionalized equation of motion for the inverted pendulum (Eq.~(\ref{eqn:inv_pend_nondim_fin})), this is an ordinary differential equation with a periodic coefficient in the zeroth-order term.


We know that for an elastic beam on a Winkler foundation with constant stiffness per unit length $K(x) = K_0$, buckling instability is triggered at a critical axial force \cite{brau2013wrinkle} 

\begin{equation}
    P_{cr} = 2\sqrt{K_0 EI},
\end{equation}

leading to the formation of a sinusoidal pattern with wavelength

\begin{equation}
    \lambda_0 = 2\pi\left(\frac{EI}{K_0}\right)^{1/4}.
\end{equation}

As in Section~\ref{sec:inverted_pendulum}, we non-dimensionalize Eq. (\ref{eq:Eq1B}) to reduce the number of independent variables from six ($E$, $I$, $P$, $K_0$, $K_1$, $\Omega$) to three. We begin by introducing the normalized coordinate 

\begin{equation}\label{barX}
   \bar{x} = \frac{x}{\lambda_0},
\end{equation}

Substitution of Eq. (\ref{barX}) into Eq. (\ref{eq:Eq1B}) yields

\begin{align}
    \frac{EI}{\lambda_0^4} y^{(4)} + \frac{P}{\lambda_0^2} y'' + \left(K_0 + K_1 \cos{\Omega \lambda_0 \bar{x}} \right) \, y &= 0\\
    EI \cdot \frac{K_0}{(2\pi)^4 EI} y^{(4)} + P \cdot \frac{\sqrt{K_0}}{(2\pi)^2\sqrt{EI}} y'' + \left(K_0 + K_1 \cos{\Omega \lambda_0 \bar{x}} \right) \, y &= 0,
\end{align}

where derivatives $y''$ and $y^{(4)}$ are taken with respect to $\bar{x}$. We now multiply both sides by $(2\pi)^4 / K_0$. This gives

\begin{align}
    \frac{(2\pi)^4}{K_0} \left[ \frac{K_0}{(2\pi)^4} y^{(4)} + P \cdot \frac{\sqrt{K_0}}{(2\pi)^2\sqrt{EI}} y'' + \left(K_0 + K_1 \cos{\Omega \lambda_0 \bar{x}} \right) \, y \right] &= \frac{(2\pi)^4}{K_0} \cdot 0 \\
    y^{(4)} + \frac{P (2\pi)^2}{\sqrt{K_0 EI}} y'' + \left(16\pi^4 + \frac{16\pi^4 K_1}{K_0} \cos{\Omega \lambda_0 \bar{x}} \right) y &= 0.  \label{eq:winkler_norm_1}
\end{align}

Eq. (\ref{eq:winkler_norm_1}) shows that the Winkler foundation’s behavior is governed by three dimensionless quantities: the normalized axial load

\begin{align}
    \bar{P} = \frac{4\pi^2 P}{\sqrt{K_0 EI}} = 8\pi^2 \frac{P}{P_{cr}}, \label{eq:winkler_bar_p}
\end{align}

the nondimensionalized amplitude of modulation stiffness 

\begin{equation} \label{eq:winkler_bar_K1}
    \bar{K} = \frac{K_1}{K_0},
\end{equation}

and the nondimensionalized frequency of modulation

\begin{equation} \label{eq:winkler_omega_bar}
    \bar{\Omega} = \Omega \lambda_0.
\end{equation}

We then substitute Eqs.~(\ref{eq:winkler_bar_p}), (\ref{eq:winkler_bar_K1}), and (\ref{eq:winkler_omega_bar}) into Eq.~(\ref{eq:winkler_norm_1}), which yields

\begin{equation}\label{eqn:winkler_nondim_fin}
    y^{(4)} + \bar{P} y'' + 16\pi^4 \left(1 + \bar{K} \cos{\bar{\Omega}\bar{x}} \right) \, y = 0,
\end{equation}

which is the non-dimensionalized equation of motion for the system about $y(x) = 0$.

Finally, we rewrite Eqn.~(\ref{eqn:winkler_nondim_fin}) as a first order system of equations, which is the form used for Floquet analysis (discussed further in Section~\ref{sec:floquet_theory}) 

\begin{equation} \label{eqn:winkler_nondim_system}
    \begin{pmatrix}
        y' \\ y'' \\ y'''\\ y^{(4)}
    \end{pmatrix} =
    \begin{pmatrix}
        0&1&0&0\\
        0&0&1&0\\
        0&0&0&1\\
        -16\pi^4 - 16\pi^4 \bar{K} \cos{\bar{\Omega}\bar{x}} & 0 & -\bar{P} & 0
    \end{pmatrix}
    \begin{pmatrix}
        y\\y'\\y''\\y'''
    \end{pmatrix}.
\end{equation}

\section{Floquet Theory}\label{sec:floquet_theory}
Equations (\ref{eqn:inv_pend_nondim_fin}) and (\ref{eqn:winkler_nondim_fin}) show that the perturbed motion of both the parametric pendulum and the modulated Winkler foundation are governed by linear differential equations with periodically varying coefficients. To analyze such equations, we use Floquet theory \cite{richards2012analysis}.
Floquet theory provides a framework for analyzing Hill equations---that is, linear differential equations of order $N$ with periodically varying coefficients of the form
\begin{equation}\label{e111}
y^{(N)} + a_{N-1}(\xi)y^{(N-1)} + \cdots + a_1(\xi) y' + a_0(\xi) y=0,
\end{equation}
where the coefficients $a_i(\xi)$ are all periodic functions with a common period $T$.
As a first step to determine our solution, we rewrite Eq. (\ref{e111}) as a first order system of equations,
\begin{equation} \label{eq:floquet_system}
    Y' = J(\xi) \, Y,
\end{equation}
where $Y \in \mathbb{R}^N$. Since all our coefficients $a_i$ are $T$-periodic, it follows that $J(\xi)$ is also $T$-periodic. Specific forms of the matrix $J$ for the parametric pendulum and the modulated Wrinkler foundation are given in Eqs. (\ref{eqn:inv_pend_nondim_system}) and (\ref{eqn:winkler_nondim_system}), respectively. 
Floquet theory tells us that solutions to Eq.~(\ref{eq:floquet_system}) take the form

\begin{equation}\label{eqsol}
    Y(\xi) = \sum_{n = 1}^N r_n(\xi) \cdot e^{s_n \xi},
\end{equation}
where $r_n(\xi) \in \mathbb{C}^N$ is periodic with period $T$ and $s_n \in \mathbb{C}$ are the Floquet exponents~\cite{lazarus2015stability, calico1984control}. 
Each $r_n(\xi) \cdot e^{s_n \xi}$ term is called a Floquet form; since $r_n(\xi)$ is periodic, the stability of each Floquet form depends entirely on the $e^{s_n \xi}$ term, or more particularly, the real component of $s_n$.
Depending on the initial or boundary conditions of Eq.~(\ref{eq:floquet_system}), $s_n$ with real parts greater than 0 could lead to growing or unbounded $Y(\xi)$ as $\xi \rightarrow \infty$.
In this work, we generally only calculate the Floquet exponents to understand the behavior of our system rather than calculate the whole solution. 
As a note, Floquet exponents are not unique; since $e^{ia} = e^{i(a + 2\pi)}$ for $a \in \mathbb{R}$, the imaginary parts of our $s_n$ are not unique.
This means that if $s = a + bi$ is a valid Floquet exponent, so is $s = a + (b + 2\pi k/T) i$ for $k \in \mathbb{Z}$. 

There are a few methods to calculate the Floquet exponents, for example, the Monodromy matrix method \cite{lazarus2015stability, calico1984control} or Hill (spectral) method \cite{hill1886part, deconinck2007spectruw, bentvelsen2017modal}. 
Here, we use the Hill method. 
In this method, we transform both our matrix $J(\xi)$ and states $Y$ and $Y'$ into the frequency space. More specifically, $J(\xi)$ and $Y(\xi)$ transform into

\begin{equation}\label{eq:floquet_J_trans}
    J(\xi) = J_0 e^{0 i \Omega \xi} + J_1 e^{i \Omega \xi} + J_{-1} e^{-i \Omega \xi} + J_2 e^{2i \Omega \xi} + J_{-2} e^{-2 i \Omega \xi} + \ldots,
\end{equation}
and
\begin{equation}\label{eq:floquet_Y_trans}
 \hat{Y}(\xi) = \sum_{h \in \mathbb{Z}} \sum_{n = 1}^N r_n^h \cdot e^{(ih\Omega + s_n) \xi},
\end{equation}
where $\Omega = 2\pi / T$ is the fundamental frequency of $J(\xi)$ and the $r_n^h$ are $N$-dimensional complex vectors that are the $h^{\text{th}}$ harmonics of the periodic function $r_n(\bar{x}) = r_n(\bar{x}+\bar{\lambda})$ of the $n^{\text{th}}$ Floquet form. 

To reduce bulk in our equations, instead of inserting the full solution into Eq.~(\ref{eq:floquet_system}), we will instead use the ansatz 

\begin{equation}\label{eq:floquet_y_ansatz}
    \hat{Y}(\xi) = \sum_{h \in \mathbb{Z}} r^h \cdot e^{(ih\Omega + s) \xi}.
\end{equation}


Substitution of Eqs.~(\ref{eq:floquet_J_trans}) and (\ref{eq:floquet_y_ansatz}) into Eq.~(\ref{eq:floquet_system}) yields

\begin{equation}\label{eq:floquet_expanded}
    0 = \sum_{h \in \mathbb{Z}} \left( J_0  + J_1 e^{i \Omega \xi} + J_{-1} e^{-i \Omega \xi} + J_2 e^{2i \Omega \xi} + J_{-2} e^{-2 i \Omega \xi} + \ldots \right)  r^h \, e^{(ih\Omega + s) \xi} - (ih\Omega + s)\, r^h \, e^{(ih\Omega + s) \xi},
\end{equation}
which can be simplified to (by factoring out the $e^{s\xi}$ term)
\begin{equation}\label{eq:floquet_expanded_no_s}
    0 = \sum_{h \in \mathbb{Z}} \left( J_0  + J_1 e^{i \Omega \xi} + J_{-1} e^{-i \Omega \xi} + J_2 e^{2i \Omega \xi} + J_{-2} e^{-2 i \Omega \xi} + \ldots \right)  r^h \, e^{ih\Omega \xi} - (ih\Omega + s)\, r^h \, e^{ih\Omega \xi}.
\end{equation}

Now, we apply the harmonic balance method: if this whole equation is balanced, each $e^{i h \Omega \xi}$ term must also be balanced. 
We show a few terms around $h = 0$ to get an idea of the general pattern.

\begin{align}
    e^{0 i \Omega \xi}: & \; 0 = J_0 r^0 - (\cancel{i 0 \Omega} + s) r^0 + J_1 r^{-1} + J_{-1} r^1 + r^2 J_{-2} + J_2 r^{-2} + \ldots\\
    e^{1 i \Omega \xi}: & \; 0 = J_0 r^1 - (i \Omega + s) r^1 + J_1 r^0 + J_{-1} r^2 + J_2 r^{-1} + J_{-2} r^3 + \ldots\\
    e^{-1 i \Omega \xi}: & \; 0 = J_0 r^{-1} - (-i \Omega + s) r^{-1} + J_1 r^2 + J_{-1} r^0 + J_2 r^{-3} + J_{-2} r^1 + \ldots\\
    e^{2 i \Omega \xi}: & \; 0 = J_0 r^2 - (2 i \Omega + s) r^2 + J_1 r^1 + J_{-1} r^3 + J_2 r^0 + J_{-2} r^4 + \ldots\\
    e^{-2 i \Omega \xi}: & \; 0 = J_0 r^{-2} - (-2 i \Omega + s) r^{-2} + J_1 r^{-3} + J_{-1} r^{-1} + J_2 r^{-4} + J_{-2} r^0 + \ldots
\end{align}


We now use these gathered terms to rewrite Eq.~(\ref{eq:floquet_expanded_no_s}) in matrix form; we get

\begin{equation} \label{eq:H_matrix}
        \left(\begin{bmatrix}
        \ddots & \vdots & \vdots & \vdots & \vdots & \vdots & \\
        \cdots & J_0 + 2i\Omega \, I_n & J_{-1} & J_{-2} & J_{-3} & J_{-4} & \cdots\\
        \cdots & J_1 & J_0 + i\Omega \, I_n & J_{-1} & J_{-2} & J_{-3} & \cdots\\
        \cdots & J_2 & J_1 & J_0 & J_{-1} & J_{-2} & \cdots\\
        \cdots & J_3 & J_2 & J_1 & J_0 - i\Omega \, I_n & J_{-1} & \cdots\\
        \cdots & J_4 & J_3 & J_2 & J_1 & J_0 - 2\Omega \, I_n & \cdots\\
        & \vdots & \vdots & \vdots & \vdots & \vdots & \ddots
    \end{bmatrix} - s\, \mathbf{I} \right)
    \begin{bmatrix}
        \vdots\\r^{-2}\\r^{-1}\\r^0\\r^1\\r^2\\ \vdots
    \end{bmatrix} = 0,
\end{equation}

or

\begin{equation}\label{eq:eigen}
    (\mathbf{H} - s \, \mathbf{I}) \, \mathbf{r} = \mathbf{0},
\end{equation}
where $\mathbf{I}$ is the identity matrix.

We see from Eq. (\ref{eq:eigen}) that Eq.~(\ref{eq:floquet_expanded_no_s}) can be rewritten as an eigenvalue-eigenvector problem of $\mathbf{H}$. 
When we retain an infinite number of terms, Eq. (\ref{eq:eigen}) is identical to Eq.~(\ref{eq:floquet_expanded_no_s}). 
Practically however, we truncate the Hill matrix to $M$ harmonics---between $r^{-M}$ to $r^M$ with $M$ between 5--9.
This leads to a system of size $(2M + 1)N$, where $N$ is the order of the considered differential equation. 
Since this truncation can introduce incorrect eigenvalues \cite{bentvelsen2017modal}, we keep only eigenvalues $s$ within the first Brillouin zone

\begin{equation}\label{eq:floquet_valid}
    |\im{s}| < \frac{\Omega}{2} + 10^{-M}, 
\end{equation}

up to a tolerance of $10^{-M}$. 
As long as we pick $M$ large enough this truncation and sorting is guaranteed to give us accurate eigenvalues \cite{zhou2004spectral}.

Below, we discuss problem-specific considerations for the inverted pendulum and the periodic Winkler foundation.

\subsection{Floquet theory for the Inverted Pendulum}

Here, we apply Floquet theory to the inverted pendulum considered in Section~\ref{sec:inverted_pendulum}. It follows from Eq. (\ref{eqn:inv_pend_nondim_system}) that the $J$ matrix for this system is given by 

\begin{equation}
    J(\tau)= \begin{pmatrix}
        0 & 1 \\ -1 - \bar{A}\cos{\bar{\Omega}\tau} & -\bar{C}
    \end{pmatrix}.
\end{equation}

Following the procedure described in the above section, we rewrite $J(\tau)$ as a Fourier expansion to obtain

\begin{equation}\label{eq:inv_pend_system_exp}
    \begin{pmatrix}
        0 & 1 \\ -1 - \bar{A}\cos{\bar{\Omega}\tau} & -\bar{C}
    \end{pmatrix} = 
    \underbrace{\begin{pmatrix}
        0 & 1\\ -1 & -\bar{C}
    \end{pmatrix}}_{J_0} \, 1 +
    \underbrace{\begin{pmatrix}
        0 & 0\\
        -\bar{A}/2 & 0
    \end{pmatrix}}_{J_1} \, e^{i\bar{\Omega}\tau} +
    \underbrace{\begin{pmatrix}
        0 & 0\\
        -\bar{A}/2 & 0
    \end{pmatrix}}_{J_{-1}} \, e^{-i\bar{\Omega}\tau}.
\end{equation}

Note that $J_1 = J_{-1}$ due to the cosine in $J(\tau)$. 
Using the $J_i$ defined in Eq. (\ref{eq:inv_pend_system_exp}), we can construct our $\mathbf{H}$ matrix using $M = 9$ as described in Eq.~(\ref{eq:H_matrix}) and then calculate its eigenvalues.
While this gives $(2M + 1)N = 38$ eigenvalues, we keep only those which satisfy Eq.~(\ref{eq:floquet_valid}).

\begin{figure}
    \centering
    \includegraphics{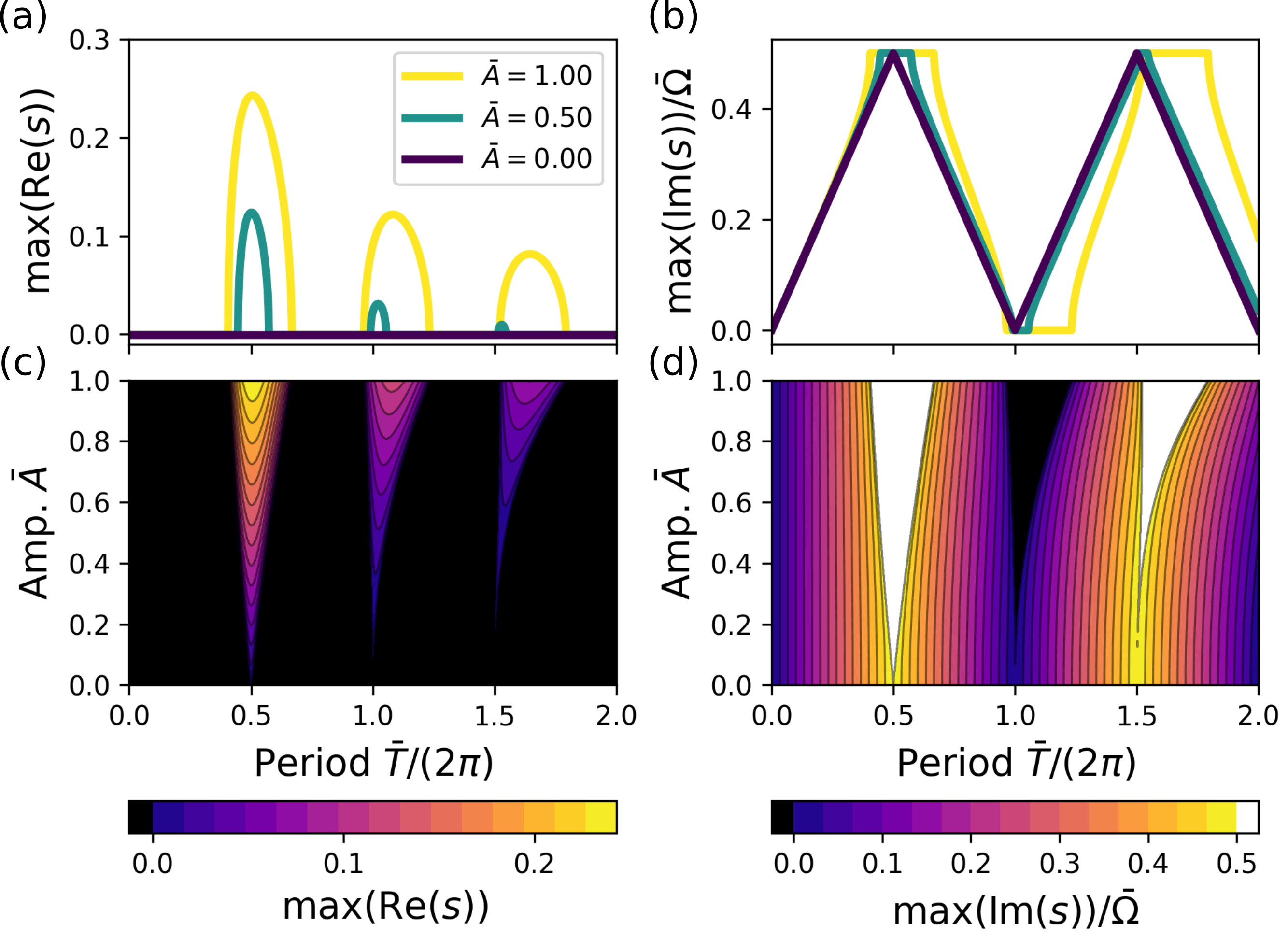}
    \caption{\textbf{Floquet exponents for the inverted pendulum} Maximum real (a) and imaginary (b) components of Floquet exponents within the first Brillouin zone for our inverted pendulum problem for a few values of non-dimensionalized amplitude, $\bar{A}$, as a function of non-dimensionalized period, $\bar{T}/(2\pi)$. (c) and (d) are max(Re(s) and max(Im(s)) in the modulation parameter space $\bar{A}$ and $\bar{T}/(2\pi)$; we can clearly see the emergence of stability tongues which are known to exist for this problem.}
    \label{fig:si_inv_pen_floquet}
\end{figure}

In Fig.~\ref{fig:si_inv_pen_floquet} we present results for an inverted pendulum with dimensionless damping factor $\bar{C} = 0.001$. Specifically, in Fig.~\ref{fig:si_inv_pen_floquet}a-b, we report the evolution as a function of $\bar{T}/(2\pi)$ of the maximum real and imaginary components of all calculated Floquet exponents in the first Brillouin zone, $\max(\Re(s))$ and $\max(\Im(s))$, for a few values of normalized amplitude $\bar{A}$. Further, in Fig.~\ref{fig:si_inv_pen_floquet}c-d, we report the evolution of $\max(\Re(s))$ and $\max(\Im(s))$ as a function of $\bar{T}/(2\pi)$ and $\bar{A}$. 

Two key features emerge from these plots. 
First, for certain ranges of $\bar{T}$, the maximum real part of the Floquet exponents, $\max(\Re(s))$, is positive. 
Since this is an initial value problem, the presence of any Floquet exponent with a positive real part indicates an exponentially growing mode in time, implying that the trivial solution $\theta(\tau) = 0$ solution is unstable. 
When we plot $\max(\Re(s))$ in the full $\bar{A}$--$\bar{T}$ parameter space, we see the emergence of instability tongues (Fig.~\ref{fig:si_inv_pen_floquet}c); these are called parametric instability regions in a dynamic system like the pendulum.
Second, in the regimes where the solution is unstable, we have either $\max(\Im(s)) = 0$ or $\max(\Im(s)) / \bar{\Omega} = 0.5$. While the real parts of the Floquet exponents determine the stability of the system, the imaginary parts provide insight into the spectral characteristics of the response.
We recall the Fourier transform of our solution (Eq.~(\ref{eq:floquet_y_ansatz})) modified to account for viscous damping \cite{lazarus2015stability}

\begin{equation}\label{eq:1234}
    \hat{\theta}(\tau) = \sum_{h \in \mathbb{Z}} r^h \cdot e^{(ih\bar{\Omega} + s - \bar{C}/2) \tau} + \sum_{h \in \mathbb{Z}} r^{*h} \cdot e^{(ih\bar{\Omega} -s - \bar{C}/2) \tau},
\end{equation}
where $ r^{*h}$ denotes the complex conjugate of $r^{h}$.
From Eq.~(\ref{eq:1234}), we can see that when $+\Im(s)$ and $-\Im(s)$ are separated by an integer multiple of $\bar{\Omega}$, the frequency spectrum contains frequencies that are integer multiples of a fundamental frequency, so our solution is periodic.
More specifically, when $\Im(s)/\bar{\Omega} = 0$, $\theta$ is periodic with period $\bar{T}$, while when $\Im(s)/\bar{\Omega} = 0.5$, $\theta$ is periodic with period $2\bar{T}$.
This phenomenon is called ``frequency lock-in'', as the output frequency of the solution locks into a half or a full fundamental frequency of the driving frequency.
Due to the structure of our problem, when we have frequency lock-in, $\max(\Re(s)-\bar{C})$ is positive and the solution grows unbounded with time.
By contrast, when the imaginary components $\pm \Im(s)$ of our Floquet exponents are not separated by an integer multiple of $\bar{\Omega}$, i.e., $\Im(s)/\bar{\Omega} \in (0, 0.5)$, the frequency spectrum of the solution contains frequencies that are incommensurate (non-integer ratios); the spectrum includes combinations of base frequencies without a single common period, so our solution is quasi-periodic. In these regions, $\Re(s) = -\bar{C}/2$, so the solution is also underdamped.

\subsection{Floquet theory for the Periodic Winkler Foundation}

Here, we apply Floquet theory to the periodic Winkler foundation considered in Section.~\ref{sec:inverted_pendulum}. 
It follows from Eq. (\ref{eqn:winkler_nondim_system}) that the
$J$ matrix for this system is given by 

\begin{equation}
   \mathbf{J}(\bar{x}) =
    \begin{pmatrix}
        0&1&0&0\\
        0&0&1&0\\
        0&0&0&1\\
        -16\pi^4 - 16\pi^4 \bar{K} \cos{\bar{\Omega}\bar{x}} & 0 & -\bar{P} & 0
    \end{pmatrix}.
\end{equation}

Following the procedure described in the above section, we first rewrite $J(\bar{x})$ as a Fourier expansion to obtain

\begin{equation}
    \begin{pmatrix}
        0&1&0&0\\
        0&0&1&0\\
        0&0&0&1\\
        -16\pi^4 - 16\pi^4 \bar{K} \cos{\bar{\Omega}\bar{x}} & 0 & -\bar{P} & 0
    \end{pmatrix} =
    \underbrace{\begin{pmatrix}
        0&1&0&0\\
        0&0&1&0\\
        0&0&0&1\\
        -16\pi^4 & 0 & -\bar{P} & 0
    \end{pmatrix}}_{J_0} \, 1 +
    \underbrace{\begin{pmatrix}
        0&0&0&0\\0&0&0&0\\0&0&0&0\\
        -8\pi^2 \bar{K} &0&0&0
    \end{pmatrix}}_{J_1} \, e^{i\bar{\Omega}\bar{x}} +
    \underbrace{\begin{pmatrix}
        0&0&0&0\\0&0&0&0\\0&0&0&0\\
        -8\pi^2 \bar{K} &0&0&0
    \end{pmatrix}}_{J_{-1}} \, e^{-i\bar{\Omega}\bar{x}}.
\end{equation}

As with the inverted pendulum, we note that $J_1 = J_{-1}$ due to the cosine in the driving term. 
Our parameters of interest are $\bar{K}$ and $\bar{\lambda} = 2\pi / \bar{\Omega}$ and
we choose to investigate $\bar{\lambda} \in [0.01, 2]$ and $\bar{K} \in [0, 0.5]$.

\begin{figure}
    \centering
    \includegraphics{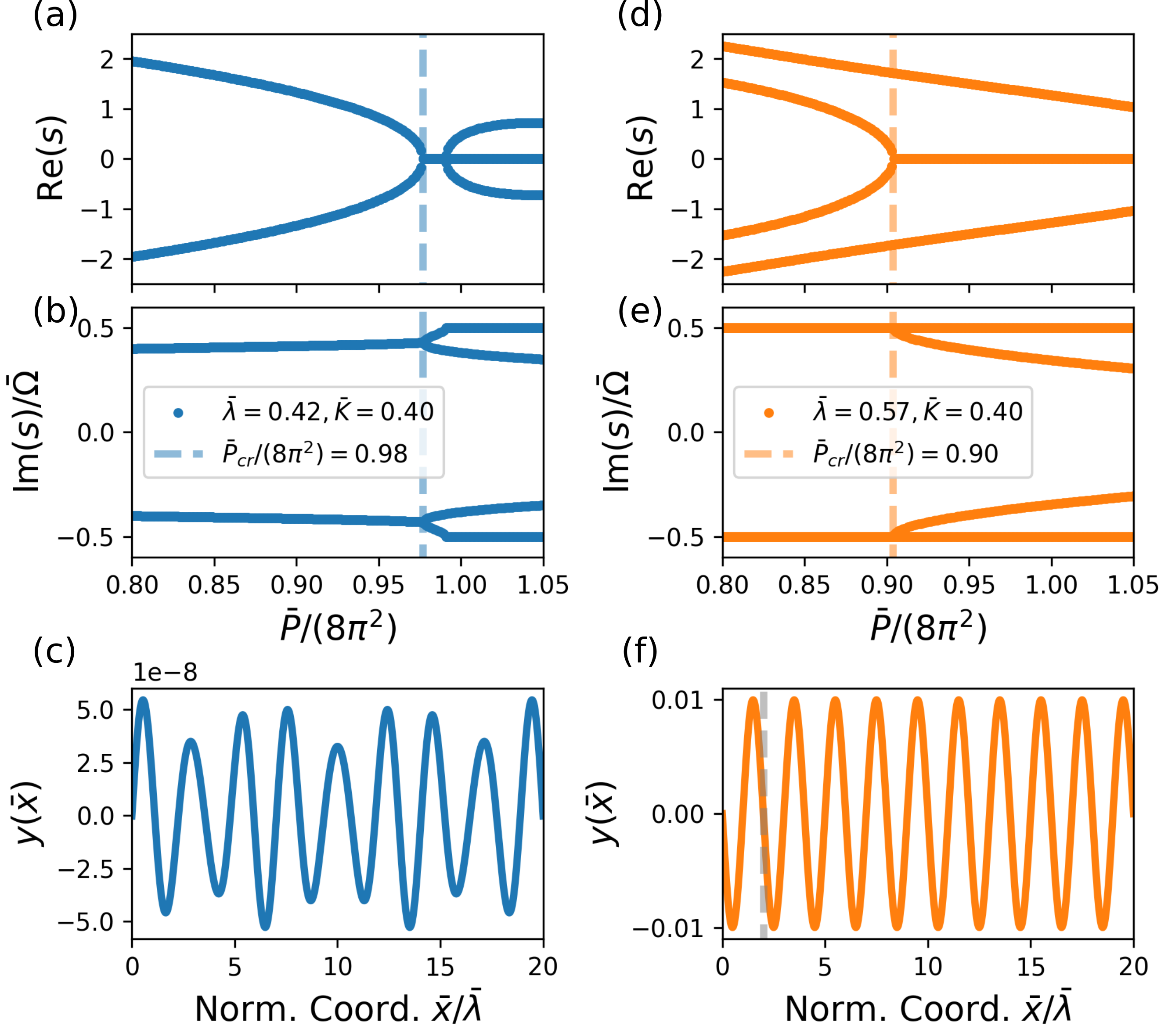}
    \caption{\textbf{Floquet exponents as a function of loading.} $\Re(s)$ (a, d), $\Im(s)$ (b, e) as a function of axial loading $\bar{P}$ and output curves (c, f) for two values of period $\bar{\lambda}$, both at $\bar{K} = 0.4$. We see the critical $\bar{P}_{cr}$ occurs when a pair of $s_i$ collapse to have real part equal to 0; this point is notated with a dashed line. The imaginary part of the Floquet exponents at the critical loading, $\bar{P}_{cr}$, tells us if our solutions is periodic (right side) or quasi-periodic (left side). The periodicity of the locked-in sample (f) is shown with a gray vertical line.}
    \label{fig:si_winkler_example_P_curves}
\end{figure}

While in initial value problems, we must have all $s$ with real component strictly less than 0 to have an asymptotically stable solution, our interpretation of Floquet exponents in the setting of boundary value problems is different.
With boundary conditions, any Floquet form $r(\xi) \cdot e^{s \xi}$ with $\Re(s) = 0$ will allow for a non trivial bounded solution, even if other Floquet forms have $\Re(s)>0$, since the coefficients in front of those Floquet forms can become zero. 
Thus, in the Winkler foundation problem, we first calculate the minimal $\bar{P}$, called the critical buckling load, $\bar{P}_{cr}$, for which at least one Floquet exponent $s$ has a real part equal to zero. In practice, to determine $\bar{P}_{cr}$, we compute the Floquet exponent $s$ for increasing values of $\bar{P}$ and identify the load at which the real part of $s$ falls below a threshold value, $\epsilon = 10^{-8}$. In Figs.~\ref{fig:si_winkler_example_P_curves}a and ~\ref{fig:si_winkler_example_P_curves}d we report the evolution of $\Re(s)$ as a function of $\bar{P}$ for two foundations with $(\bar{K}, \bar{\lambda})=(0.4, 0.42)$ and $(0.4, 0.57)$, respectively. For this two geometries we find that $\bar{P}_{cr}=0.98\cdot 8\pi^2$ and $0.90\cdot 8\pi^2$, respectively.

\begin{figure}
    \centering
    \includegraphics{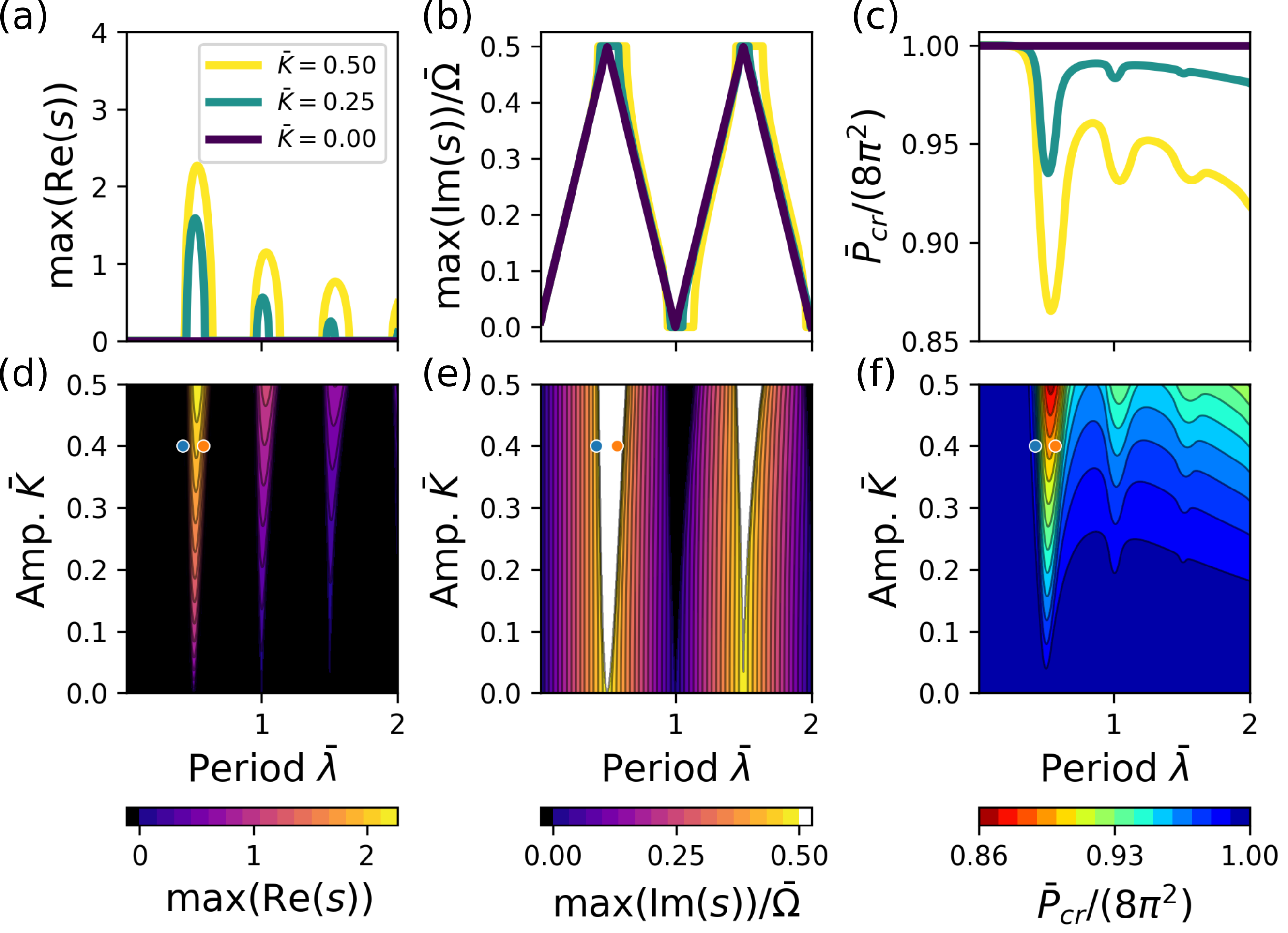}
    \caption{\textbf{Real, imaginary components of $s_n$ and the calculated critical applied load associated with each geometry.} In (a, b, c) we show the (a) maximum real component, (b) maximum imaginary component, and (c) calculated critical load as a function of $\bar{\lambda}$ for a few $\bar{K}$. In (d, e, f) we generalize these results in heat maps. Like for the inverted pendulum, we see the emergence of tongues where the maximum real component is positive. Again we see that in those tongues, the normalized imaginary component is either $0$ or $0.5$; we also see that in these tongues, there is a sharp decrease in necessary buckling force. In (d/e/f) we also plot the specific points seen in Fig.~\ref{fig:si_winkler_example_P_curves} in orange and blue.}
    \label{fig:si_winkler_floquet_proc}
\end{figure}

Once the critical buckling load for a given geometry is determined, the imaginary parts of the Floquet exponents provide insight into the spectral content of the emerging buckling mode. 
In Figs.~\ref{fig:si_winkler_floquet_proc}b and \ref{fig:si_winkler_floquet_proc}e, we plot the evolution of $\max(\Im(s)) / \bar{\Omega}$ as a function of $\bar{P}$ for the same two geometries. 
For the first case ($\bar{\lambda} = 0.42$), at $\bar{P}_{cr}$ we find that $\max(\Im(s)) = 0.43$, indicating a quasi-periodic buckling mode (see Fig.~\ref{fig:si_winkler_floquet_proc}c). 
In contrast, for the second case ($\bar{\lambda} = 0.57$), we have that $\max(\Im(s))/\bar{\Omega} = 0.5$ at $\bar{P}_{cr}$, corresponding to a frequency lock-in---similar to what is observed in the inverted pendulum---which results in a $2\bar{\lambda}$-periodic buckling mode (Fig.~\ref{fig:si_winkler_floquet_proc}f).
Finally, we note that, unlike the dynamic case, where periodic solutions grow unbounded in time and quasi-periodic ones decay (assuming positive damping), here both quasi-periodic and periodic solutions remain bounded and stable in the spatial variable $\bar{x}$.

Since the buckling load is also a function of our parameters, we plot $\bar{P}_{cr}$ normalized by the $8 \pi^2$, the critical buckling force in the unmodulated case, i.e. when $\bar{K} = 0$. These example curves for different modulation amplitudes can be seen in Fig.~\ref{fig:si_winkler_floquet_proc}c.
Again, if we plot $\max(\Re(s))$ and $\max(\Im(s))$ as a function of amplitude, $\bar{K}$, and period of modulation, $\bar{\lambda}$, we see the emergence of lock-in tongues; these can be seen in Figs.~\ref{fig:si_winkler_floquet_proc}d-f.
We see that these lock-in tongues alternate between $\Im(s) = 0$, indicating $\bar{\lambda}$-periodic solutions, and $\Im(s) = 0.5$, indicating $2\bar{\lambda}$-periodic solutions. 
We also observe that the critical force, $\bar{P}_{cr}$, needed to buckle the structure decreases rapidly at the onset of of each lock-in tongue (Fig.~\ref{fig:si_winkler_floquet_proc}f).

\clearpage
\section{Finite Element simulations}
We performed Finite Element (FE) simulations of the modulated thin strip using the commercial software packages Abaqus 2019/Standard and 2021/Standard. Two types of simulations were carried out: (1) finite-size simulations, using either a buckling step or modeling the post-buckling response, and (2) Bloch wave simulations conducted on a single periodic unit cell.

All models were meshed with shell elements, S3 or S4R, with an out-of-plane Poisson's ratio specified as $\nu = 0.49$. This shell-specific Poisson's ratio controls the change in thickness due to in-plane membrane strain.
For the material, we use an incompressible Neo-hookean material model.

\subsection{Finite size simulations} \label{sec:fea_finite}
We simulate finite size samples over a variety of lengths, ranging from $L = 200$ mm (matching the experimental samples) to $L = 1200$ mm.
Since the period of modulation, $\lammod$ will not always divide evenly into the total length, some simulations have a non-integer number of periods; this did not affect our results.
To impose uniform compression along the bottom edge, we set our boundary condition by constraining the $x$-displacement for every point along the bottom edge.
We conducted two types of finite size simulations: (1) linear buckling analyses and (2) post-buckling analyses, using one of the eigenmodes from the linear buckling analysis as a geometric imperfection.

\subsubsection{Linear buckling analyses}
\begin{figure}[h]
    \centering
    \includegraphics{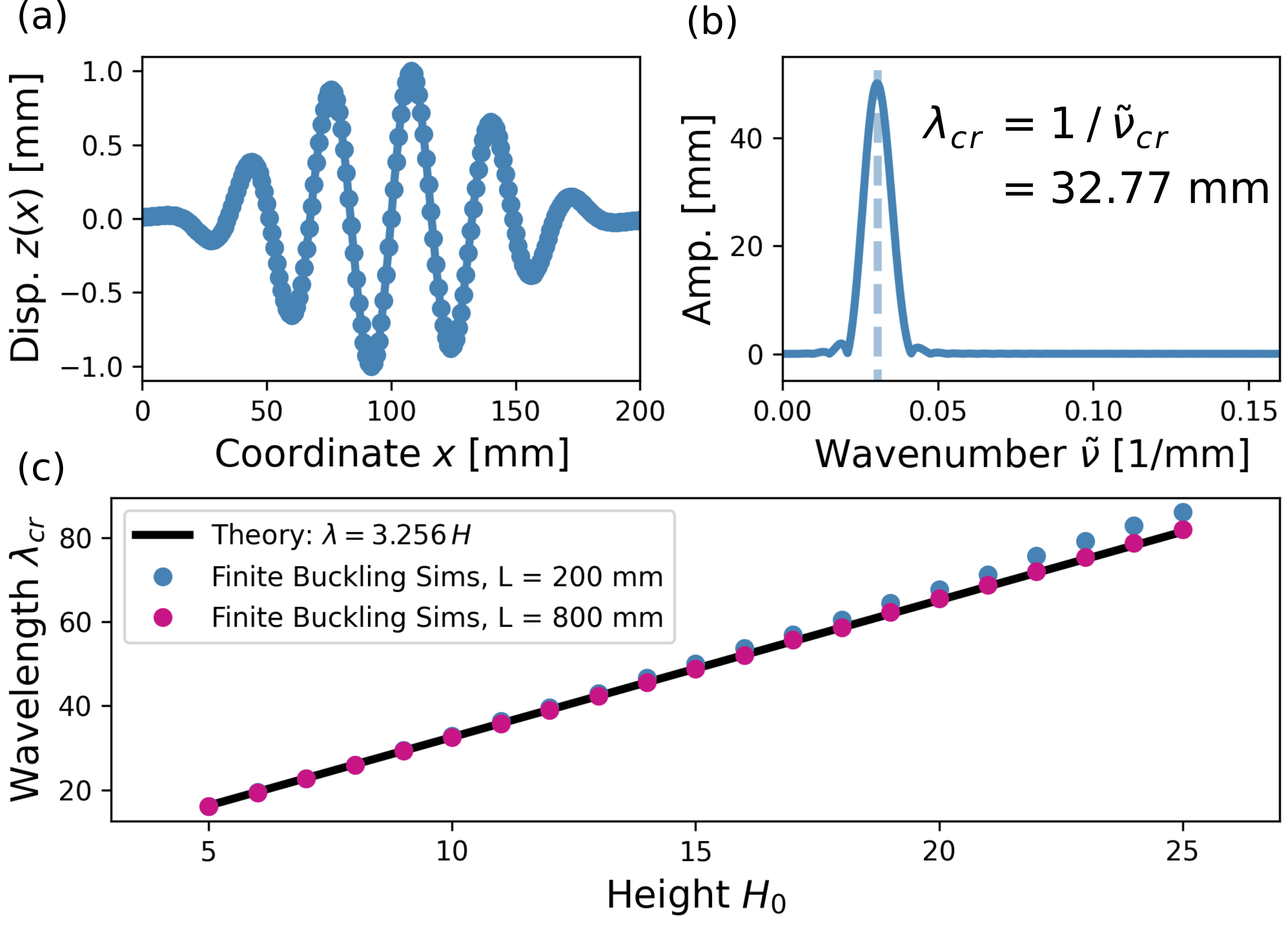}
    \caption{\textbf{Validation of unmodulated theory for finite length simulations} (a) Deformation of the first linear buckling mode for $H_0 = 10$ mm (b) DFT of that mode; we locate the wavenumber associated with the peak and calculate the wavelength of the mode. (c) Wavelength as a function of $H_0$; we see that our FE simulations match theory.}
    \label{fig:si_fea_rec}
\end{figure}

From each linear buckling analysis, we extract the $z$-displacement as a function of original $x$-coordinate of the top edge of our strip (Fig.~\ref{fig:si_fea_rec}a).
We seek to extract spectral information about the displacement using the discrete Fourier transform (DFT).
In the modulated case, our mesh is not guaranteed to be uniformly spaced in $x$, so we first linearly interpolate the $z$-displacement with 100 points per period.
We then take the DFT over our function $z(x)$ and find the wavenumber $\tilde{\nu}$ associated with the largest peak magnitude (Fig.~\ref{fig:si_fea_rec}b).
To ensure the binning is not an issue, we zero-pad our data to have a total length of $n_{\text{fft}} = 2^{14}$.

\paragraph{Unmodulated case: $H_1 = 0$.}
We first validate our simulations in the unmodulated case, where analytical work has shown that for thin strips of height $H_0$ subject to compressive loads along their bottom edge, the wavelength of the top edge should be~\cite{mora2006buckling}

\begin{equation}\label{eq:si_wave_rec}
    \lambda = 3.256 H_0.
\end{equation}

We compare this analytical result to our simulations by calculating $\lambda_{cr} = 1/\tilde{\nu}$ (Fig.~\ref{fig:si_fea_rec}b). 
This comparison can be seen in Fig.~\ref{fig:si_fea_rec}c, and we see good results.
We also observe that as our finite length $L$ increases, we see better matching with the theory, since the number of periods $L / \lammod$ increases. 
This motivates our usage of periodic unit cell simulations, discussed below, as we would like to calculate the critical wavenumbers for our system without having to use a very large $L$.

\paragraph{Modulated case: $H_1 > 0$.}

In the modulated case, we still look for the location of the wavenumber associated with the largest peak.
This gives us a single wavenumber; however, we can calculate the all valid wavenumbers in the reciprocal space via

\begin{equation}
    \tilde{\nu}^* = \frac{n}{\lammod} \pm \tilde{\nu} \; \; \text{for} \; \; n \in \mathbb{Z}.
\end{equation}
This equation comes from the Bloch wave conditions and is equivalent to our Floquet exponents $s_i$ being separated by $h\bar{\Omega}$ in our earlier analysis.
While finite size simulations are an approximation of the infinite solution, as long as our sample is long enough, we can assume this equation holds.
In the same way as the Winkler foundation, we see periodic solutions when our normalized wavenumber, $\tilde{\nu} \lammod$, has a value of $0.5 + k$ or $0 + k$ for $k \in \mathbb{Z}$. 
Otherwise, our solutions are quasi-periodic.
We plot the locations of our peaks (black) and our extended spectra (gray) as a function of $\lammod$ at $H_1 / H_0 = 0.5$ in Fig.~\ref{fig:si_fea_finite}c.

\subsubsection{Post-buckling analyses}
Besides linear buckling analyses, we also perform static analyses into the postbuckling regime.
The main purpose of these is as a closer comparison to our experiments, where we measure in the postbuckled state and where the total length of our sample is a limiting factor.
All models used in our post-buckling simulations have $L = 200$ mm, to match with experimental samples and were compressed to 10 mm (except the samples for $\lammod = 8.2$ and $12.2$ mm, which were compressed to 13 and 12 mm respectively) to match experiments.
We additionally use these analyses to check that performing our analysis in the post-buckling regime does not change the location of the peak wavenumber, $\tilde{\nu}_{cr}$.

To ensure that the structure buckles upon compression, we introduce an imperfection into the mesh in the form of the most critical eigenmode. The mesh is perturbed by the first eigenmode, $\mathbf{v}_1$, scaled by the
scale factor $\epsilon$ 

\begin{equation}
    \mathbf{X}^{\text{postbuckling}} = \mathbf{X}^{\text{original}} + \epsilon \mathbf{v}_1,
\end{equation}

where $\mathbf{X}^{\text{postbuckling}}$ and $\mathbf{X}^{\text{original}}$ are the modified and original coordinates of our model, respectively and $\epsilon$ is some scalar parameter that controls how much we edit our initial mesh; in this work we use $\epsilon = 0.05 t$, where $t$ is the thickness of our strip in the $z$-direction. 

\begin{figure}[h]
    \centering
    \includegraphics{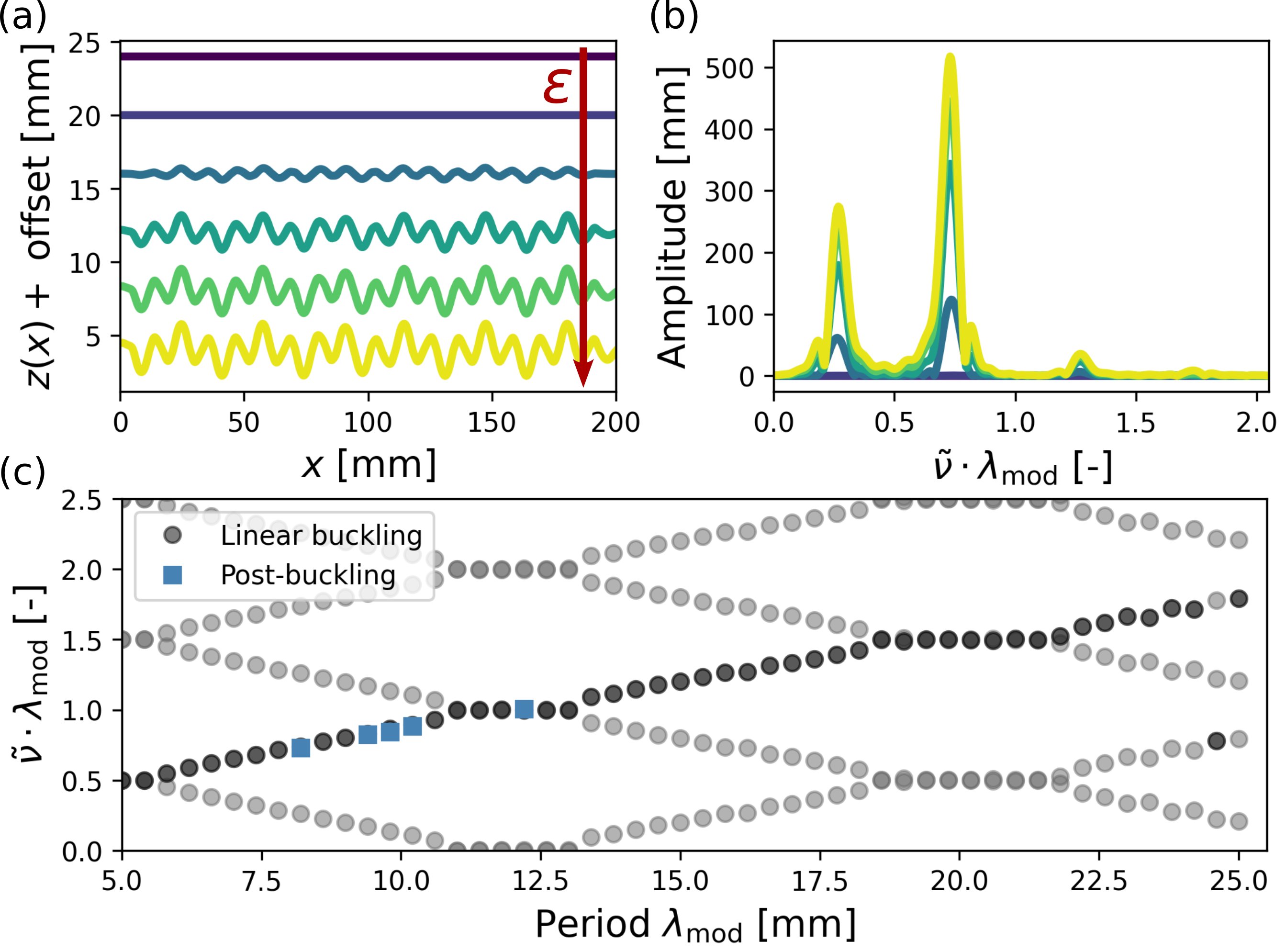}
    \caption{\textbf{Validation of postbuckling simulations} (a) Deformation of the top edge $z(x)$ for a simulation $\lammod = 8.2$ mm, $H_1 / H_0 = 0.5$ with increasing strain. (b) We see that the locations of the peaks do not significantly move in the post-buckling regime. (c) Our post-buckling simulations ($L = 200$ mm) have the same fundamental peak locations as linear buckling simulations ($L = 1200$ mm).}
    \label{fig:si_fea_finite}
\end{figure}

In Fig.~\ref{fig:si_fea_finite}a, we plot several snapshots of the top edge displacement, $z(x)$, at increasing strains for a strip with $H_1/H_0 = 0.5$ and $\lammod = 8.2$ mm.
We calculate the DFTs of each of those curves and plot those in Fig.~\ref{fig:si_fea_finite}b. We see that once peaks appear (after buckling onset), their locations do not substantially change with increasing deformation.
We then plot the location of the highest peak from the final frame of our postbuckling simulations against the extended spectra from linear buckling simulations (Fig.~\ref{fig:si_fea_finite}c); we see that the postbuckling simulations have the same peak locations as the linear buckling simulations.

\subsection{Bloch wave simulations} \label{sec:fea_bloch}
We also investigate the behavior of infinite periodic strips using simulations of a single unit cell \cite{liu2015bloch}; in the modulated case, this unit cell has length $\lammod$, while in the unmodulated case, its length is arbitrary; we denote the length of our unit cell as $L^*$.
To study the buckling of our structure, we first compress our unit cell to some strain $\epsilon$, then investigate the frequency response of small-amplitude waves of linear wavenumber $\tilde{\nu}$. 
We know the response of a wave through a system looks like \cite{brillouin1946wave}

\begin{equation}
    u(x,t) = u_0(x) \, e^{i\omega t},
\end{equation}
so if $\omega$ is real, then $u(x,t)$ describes a stable perturbation. 
However, if $\omega$ is instead imaginary, then $u(x,t)$ grows unstably.
We therefore look for the onset of this instability, when $\omega$ transitions from real to imaginary---where $\omega^2 = 0$.
The wavenumber of the buckled state and the infinite periodic solution can then be extracted from the deformed state.
To test the frequency response of a wavenumber $\tilde{\nu}$, we use Bloch wave boundary conditions along the vertical edges of our unit cell. These boundary conditions relate the deformation of the right side $\mathbf{u}^{\text{right}}$ to the left $\mathbf{u}^{\text{left}}$ as

\begin{equation}\label{eq:si_bloch_bd}
    \mathbf{u}^{\text{right}} = \mathbf{u}^{\text{left}} \, e^{i 2\pi \tilde{\nu} L^*},
\end{equation}
where $L^*$ is the undeformed length of our sample and is equal to $\lammod$ for the modulated simulations.
We pick $\tilde{\nu}$ in the first Brillouin zone, i.e. $\tilde{\nu} \in [0, 1 / L^*)$.
Practically, we know neither the buckling wavenumber $\tilde{\nu}$ nor the strain $\epsilon$ at which the structure buckles, so to find both we sweep over many values of $\epsilon$ and look for the lowest strain at which there is a 0-frequency response.

We simulate this process in ABAQUS Standard/2021, using a *STATIC step for the deformation of the unit cell up to $\epsilon$ and a *FREQUENCY step for the eigenfrequency analysis.
At each Frequency step, the boundary condition from Eq.~(\ref{eq:si_bloch_bd}) is changed to test a new value for $\tilde{\nu}$.
In this work, we used 50 Frequency steps for each geometry, which is equivalent to testing 50 values of $\tilde{\nu} \in [0, 1/L^*)$.
Additionally, because Abaqus cannot handle complex values as those required to prescibe the boundary conditions defined by Eq. (\ref{eq:si_bloch_bd}), we split our unit cell into two identical unit cells with the same mesh.
One handles the real component of the deformation, while the other handles the imaginary component of the deformation \cite{aaberg1997usage}.

\paragraph{Unmodulated case: $H_1 = 0$.}

\begin{figure}
    \centering
    \includegraphics{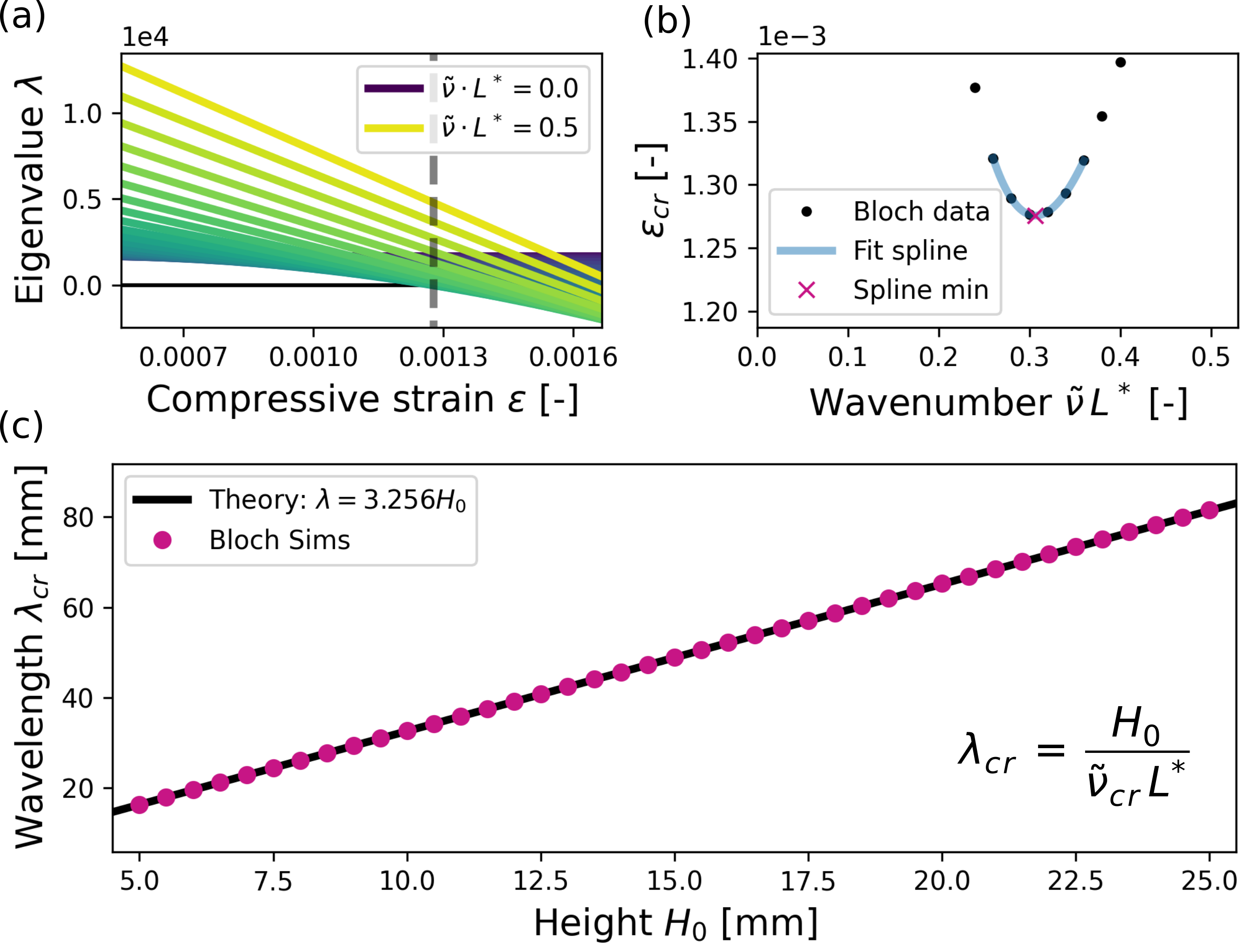}
    \caption{\textbf{Validation of unmodulated theory for Bloch wave simulations} (a) For a given geometry (here, $H_0 = 15$ mm), we sweep over many wavenumbers $\tilde{\nu}$ at different strains $\epsilon$. (b) Each wavenumber $\tilde{\nu}$ yields a 0-frequency response at a different strain, $\epsilon_{cr}$; we fit a spline to determine the minimum of the underlying function. (c) Using the calculated $\tilde{\nu}_{cr}$, we calculate the critical wavelength and compare to the theoretical result.}
    \label{fig:si_fea_bloch_rec}
\end{figure}

As with the finite sized simulations, we first seek to validate our Bloch wave simulations by checking that the theory from Eq.~(\ref{eq:si_wave_rec}) holds.
While for the unmodulated case, we can choose $L^*$ arbitrarily, we pick $L^* = H_0$, so that the expected fundamental wavenumber, $\tilde{\nu}_{cr} = 1/(3.256 \, H_0)$, lies within the lower half of the first Brillouin zone, i.e. $\tilde{\nu} L^* \in [0, 0.5]$.
We see an example of this process for $H_0 = L^* = 15$ mm in Fig.~\ref{fig:si_fea_bloch_rec}a.
Because we plot the critical wavelength, $1/\tilde{\nu}_{cr}$, we are especially sensitive to small errors in the critical wavenumber, $\tilde{\nu}_{cr}$.
Therefore, instead of simply taking the wavenumber $\tilde{\nu}_{cr}$ with the lowest critical strain, we fit a fourth order spline to the data points around that minimum and find the local minimum of that fit spline (Fig.~\ref{fig:si_fea_bloch_rec}b).
We then use that calculated $\tilde{\nu}_{cr}$ to find the critical wavelength for that given geometry; as shown in Fig.~\ref{fig:si_fea_bloch_rec}c, we find good matching with the theory.

\paragraph{Modulated case: $H_1 > 0$.}

\begin{figure}
    \centering
    \includegraphics{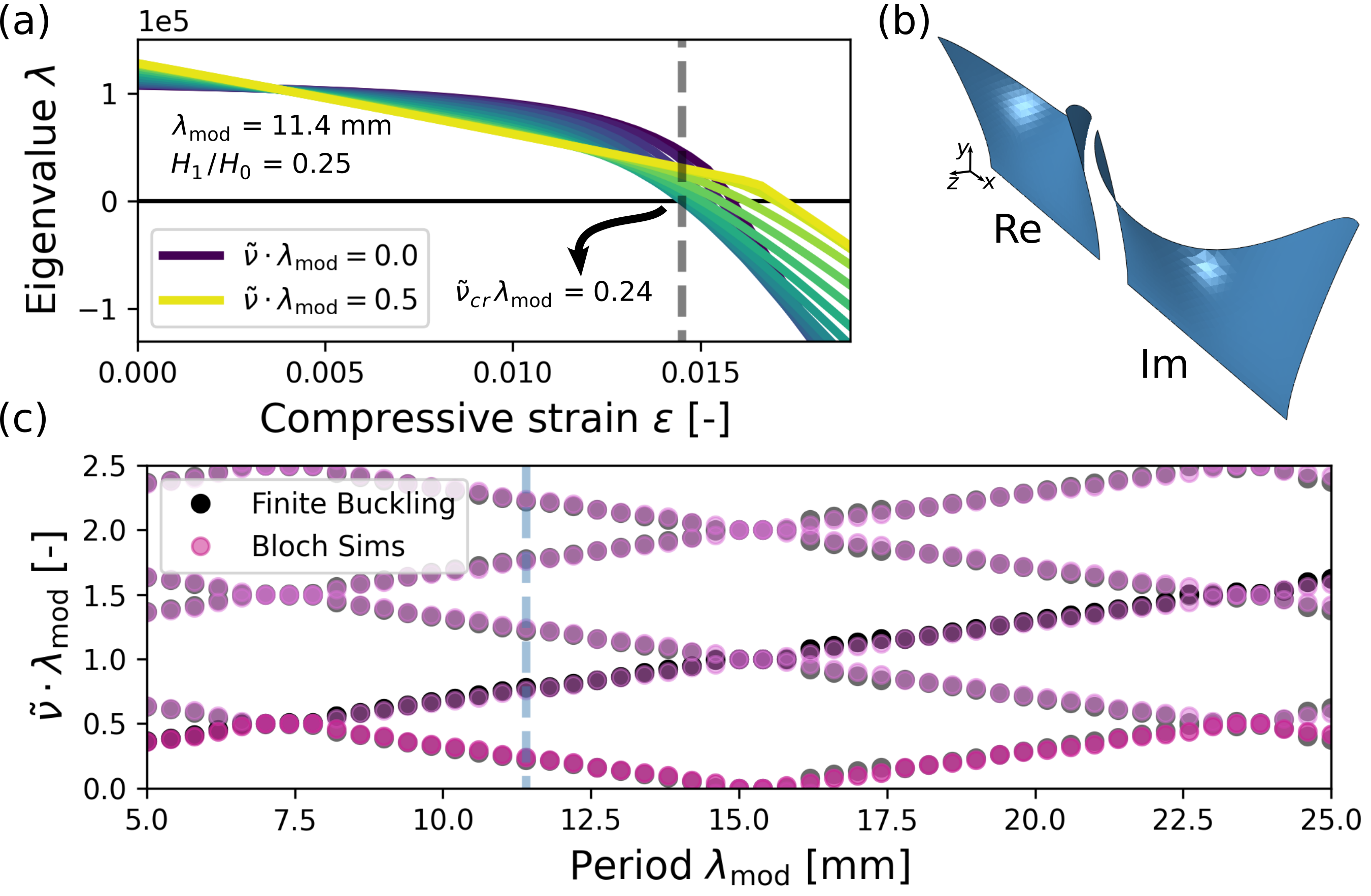}
    \caption{\textbf{Modulated Bloch wave results} (a) We search for both the critical strain $\epsilon_{cr}$ and the wavenumber $\tilde{\nu}_{cr}$ for each geometry. (b) The deformation of the complex unit cell at that critical strain; we can use its deformation to reconstruct the infinite solution. (c) Comparison of finite linear buckling simulations with $L = 1200$ mm to Bloch wave simulations for $H_1 / H_0 = 0.25$.}
    \label{fig:si_fea_bloch}
\end{figure}

We perform the same strain sweep in the modulated case, using the critical strains from finite simulations as initial guesses and only checking strain values around that guess. 
An example of this process for $\lammod = 11.4$ mm and $H_1 / H_0 = 0.25$ can be seen in Fig.~\ref{fig:si_fea_bloch}a.
As seen in the main text, we perform this process for many geometries; we see in Fig.~\ref{fig:si_fea_bloch}c that for $H_1 / H_0 = 0.25$, the extended spectra of the finite buckling and Bloch wave analysis simulations match well, validating our implementation of the periodic unit cell simulations.

Having identified the critical strain $\epsilon_{cr}$, the buckling mode of the strip can be determined from the critical eigenmode of the two meshes, defined by $\mathbf{u}^{\mathrm{Re}}$ and $\mathbf{u}^{\mathrm{Im}}$ (Fig.~\ref{fig:si_fea_bloch}b), as

\begin{equation}
    \mathbf{u}(x) = \mathbf{u}^{\mathrm{Re}}(x^*) \sin{2\pi L^* \eta} + \mathbf{u}^{\mathrm{Im}}(x^*) \cos{2\pi L^* \eta},
\end{equation}
where $x^* = x / L^*$ and $\eta = \lfloor x / L^* \rfloor$.
To calculate the fundamental wavenumber, we can reconstruct the deformation of the top edge, $z(x)$, to some large length $\ell \gg L^*$ and calculate the peak $\tilde{\nu}$ of the DFT.


\clearpage
\section{Fabrication}\label{sec:si_fabrication}
The blocks used in this study are made of nearly incompressible polyvinylsiloxane (PVS) elastomers. We use Elite Double 32 from Zhermack (with green color). The structures are fabricated as a single part using a two-step casting process. In the first step, the thin modulated strip is molded onto a thin supporting block with an out-of-plane thickness of 12.7 mm. In the second step, the block thickness is increased to 40.0 mm to apply the appropriate boundary conditions to the bottom edge of the strip. This two-step approach is necessary because direct molding of the thin modulated layer onto a thick substrate consistently led to cracking in the modulated region during demolding. Importantly, the final block thickness must be sufficiently large to prevent deformation of the block itself under compressive loading, ensuring that buckling occurs only in the modulated strip as intended.

\begin{figure*}[!hpt]
\begin{center}
\includegraphics[width = 0.89 \columnwidth]{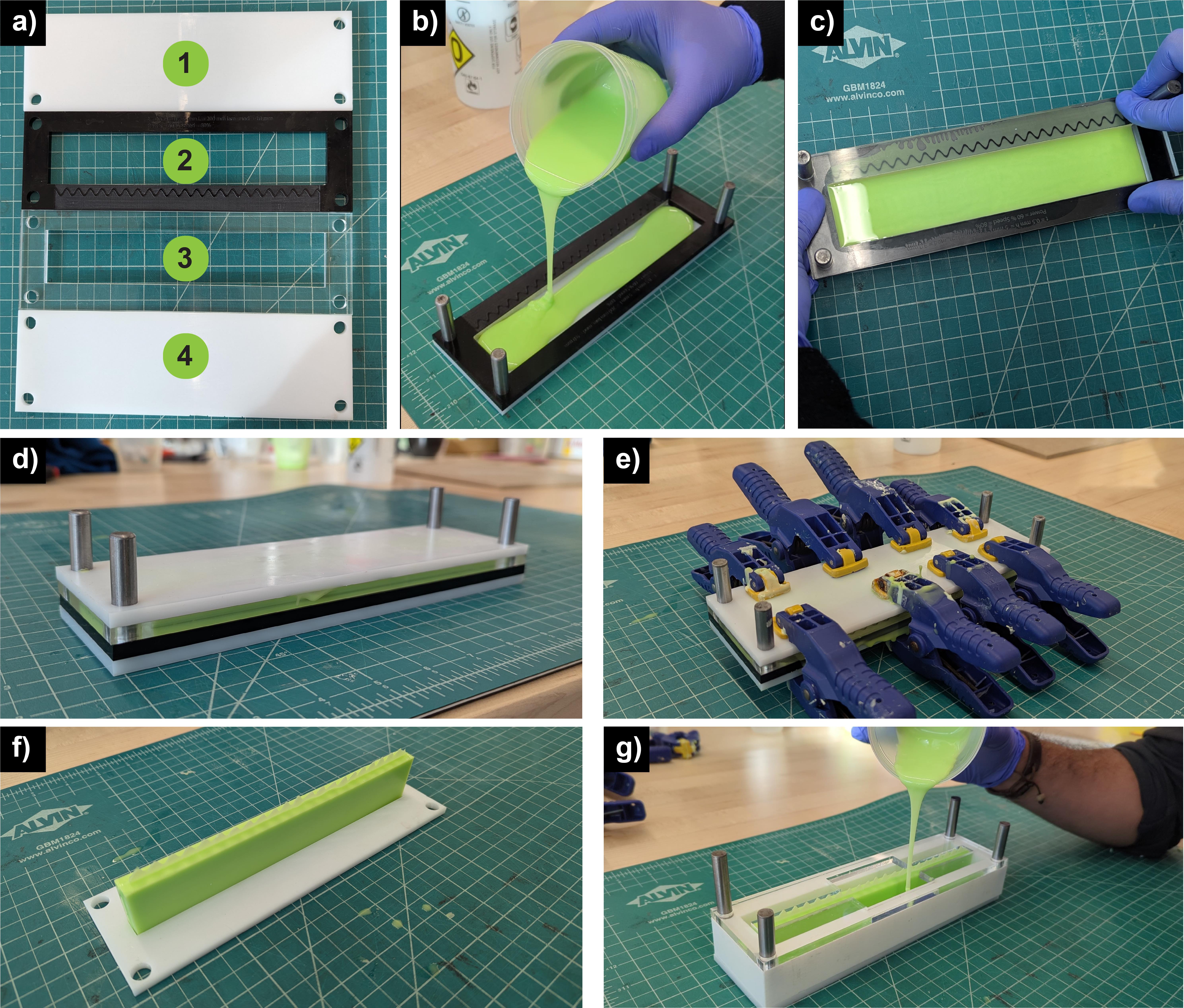} 
\caption{\label{fig:1mold} Snapshots of the main steps required to fabricate our structures.}
\vspace{-15pt}
    \end{center}
\end{figure*}

\begin{figure*}[!hpt]
\begin{center}
\includegraphics[width = 0.39 \columnwidth]{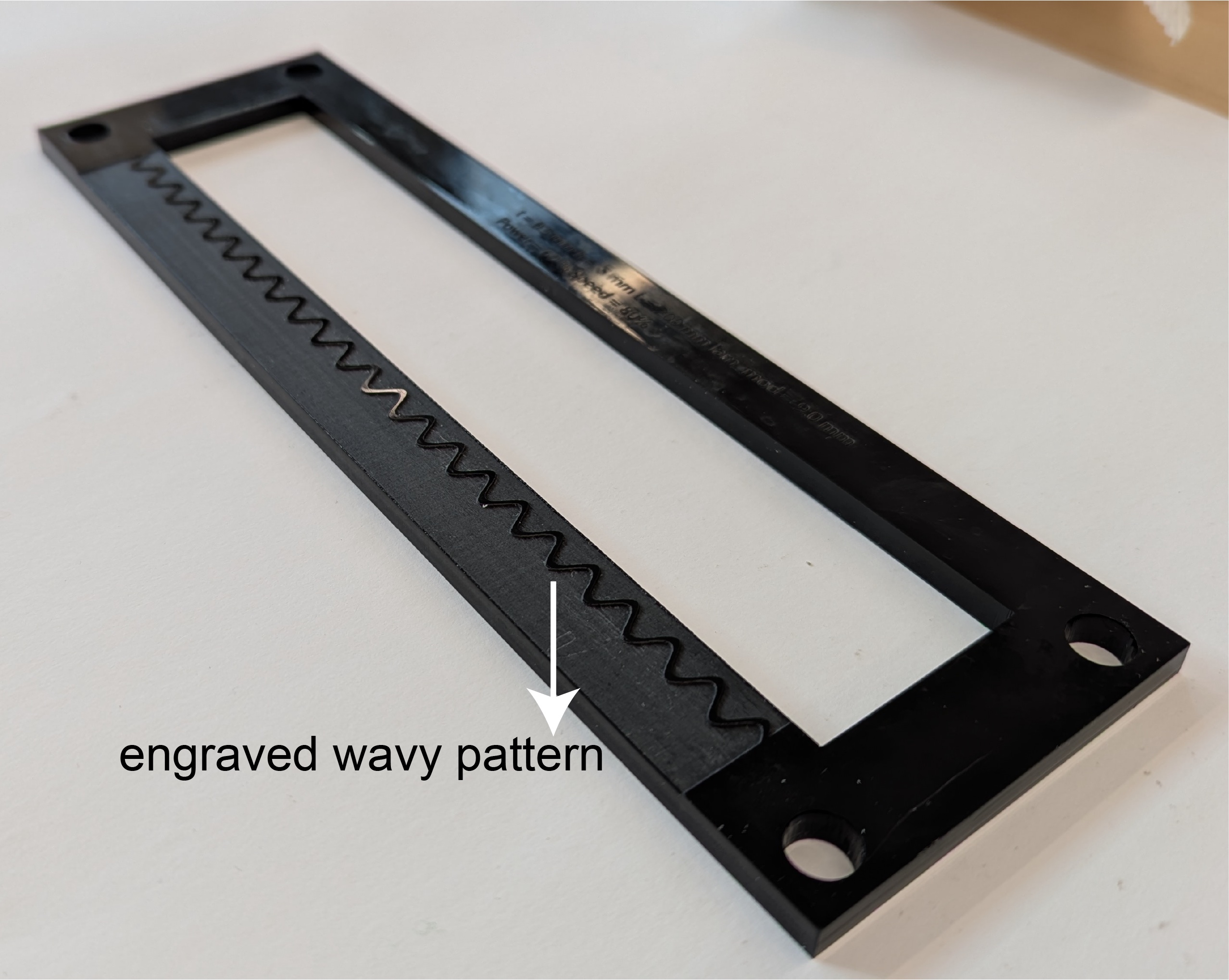} 
\caption{\label{fig:engraving} Laser cut mold with engraved waving pattern on its edge. }
\vspace{-15pt}
    \end{center}
\end{figure*}

The following step-by-step process is followed:
\begin{enumerate}
    \item \textbf{Mold Preparation for the modulated strip on a thin block:} We laser-cut four acrylic parts with 6.35 mm thickness to form the mold for the modulated strip on a thin block, as illustrated in Fig.~\ref{fig:1mold}a. The black part (part \#2) has the desired wavy pattern engraved (highlighted in Fig.~\ref{fig:engraving}). The depth of the mold can be tailored by adjusting the laser power during manufacturing. To facilitate demolding, all molds are coated with Mann Release Mold 200.

    \item \textbf{Casting the modulated strip on a thin block:} Parts \#1 and \#2 are stacked and aligned using four pins (Fig.~\ref{fig:1mold}b). The assembly is then filled with uncured PVS elastomer (Fig.~\ref{fig:1mold}c).

    \item \textbf{Closure and Curing:} Parts \#3 and \#4 (Fig.~\ref{fig:1mold}d) are placed on top of the assembly, and pressure clamps are used to ensure uniform thickness (Fig.~\ref{fig:1mold}e). The elastomer is cured at room temperature (25~$^\circ$C) for 30 minutes before demolding. At this point, the modulated strip on a thin block is ready. 

    \item \textbf{Mold Preparation for thickening the block:} A box with two open faces is 3D printed using a BambuLab X1 Carbon printer. Additionally, two top and bottom acrylic plates are laser-cut to enclose the box. 

    \item \textbf{Final casting and curing:} The cured modulated strip on a thin block is placed at the center of the bottom plate (Fig.~\ref{fig:1mold}f). The curing box is assembled, and the uncured PVS elastomer mixture is poured inside (Fig.~\ref{fig:1mold}g). The assembly is cured at room temperature (25~$^\circ$C) for 30 minutes. 

    \item \textbf{Finishing Touches:} The edges of the wavy pattern are manually colored black using a Sharpie.
\end{enumerate}


\clearpage

\section{Testing}\label{sec:si_testing}
\begin{figure}[h]
    \centering
    \includegraphics[width = 0.4 \columnwidth]{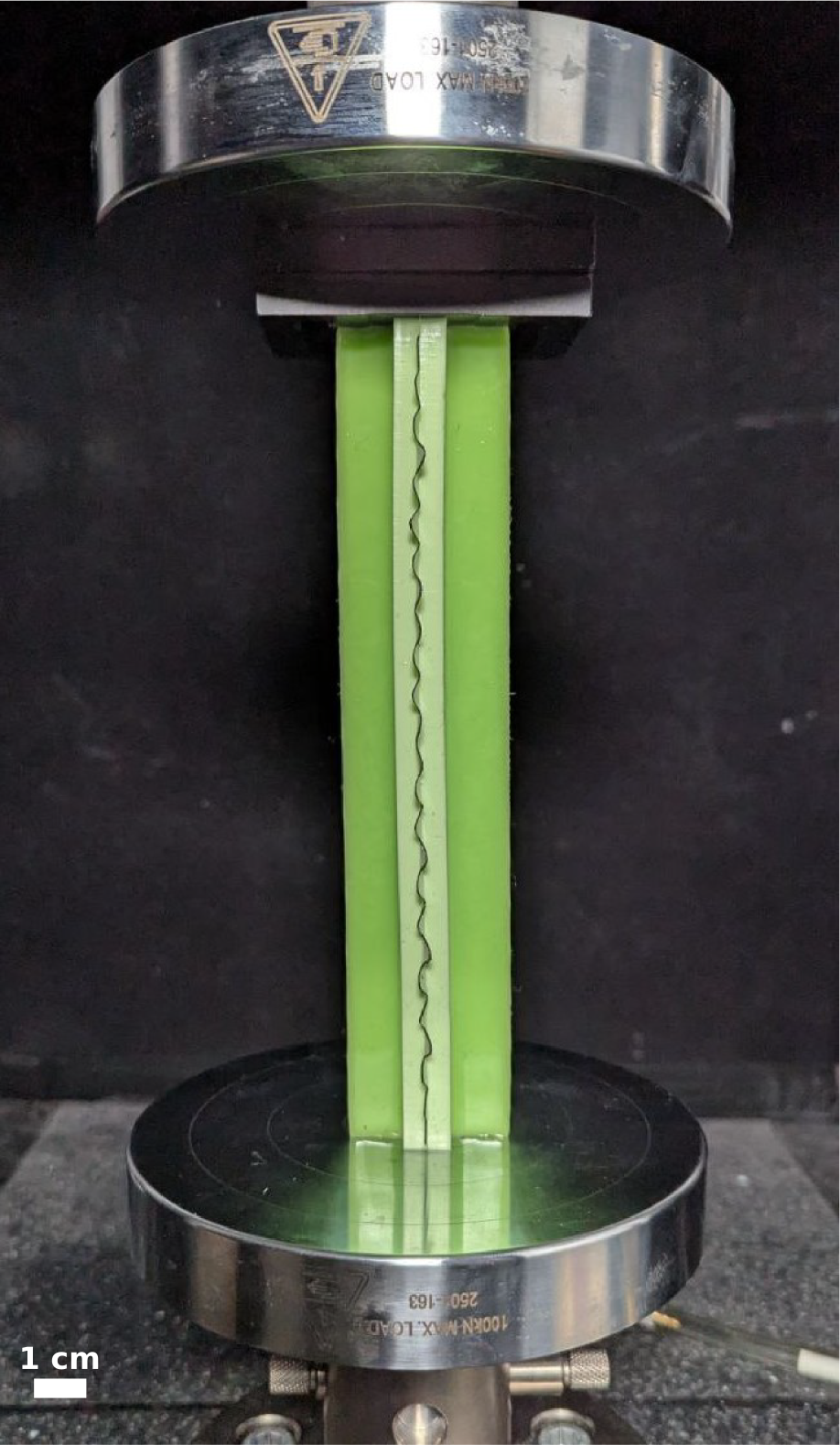} 
    \caption{\label{fig:test} Example of compression test of a structure with $\lammod = 8.2$ mm.}
\end{figure}

Our samples are compressed using a universal testing machine (Instron 5969, outfitted with a 2530-series 500-N load cell). To compress the samples uniformly, we place them within the flat circular clamps (Fig.~\ref{fig:test}). 
The deformations of the painted edge are captured with a camera and post-processed with the opencv library in Python.

In order to make the FFTs shown in Fig.~3 of the main text, we need to calculate the top displacement, $z(x)$ as a function of the original $x$ coordinate.
To calculate $z(x)$, we
\begin{enumerate}
    \item Post-process the undeformed sample to calculate the location of the top edge in the reference configuration. We define these locations to be $(x, z_0(x))$.
    \item Post-process the deformed edge to calculate the location of the top edge in the current configuration. We define this to be $(x_{\text{final}}, z_0(x_{\text{final}}))$.
    \item Using the known global deformation, $u_x$, project the deformed coordinates $(x_{\text{final}}, z_{\text{deformed}}(x_{\text{final}}))$ back into the reference frame to calculate $(x, z_{\text{deformed}}(x))$.
    \item Calculate $z(x) = z_{\text{deformed}}(x)) - z_0(x_{\text{final}}))$
    \item Calculate the DFT of $z(x)$ with $n_{\text{fft}} = 2^{14}$
\end{enumerate}

\begin{figure}[h]
    \centering
    \includegraphics{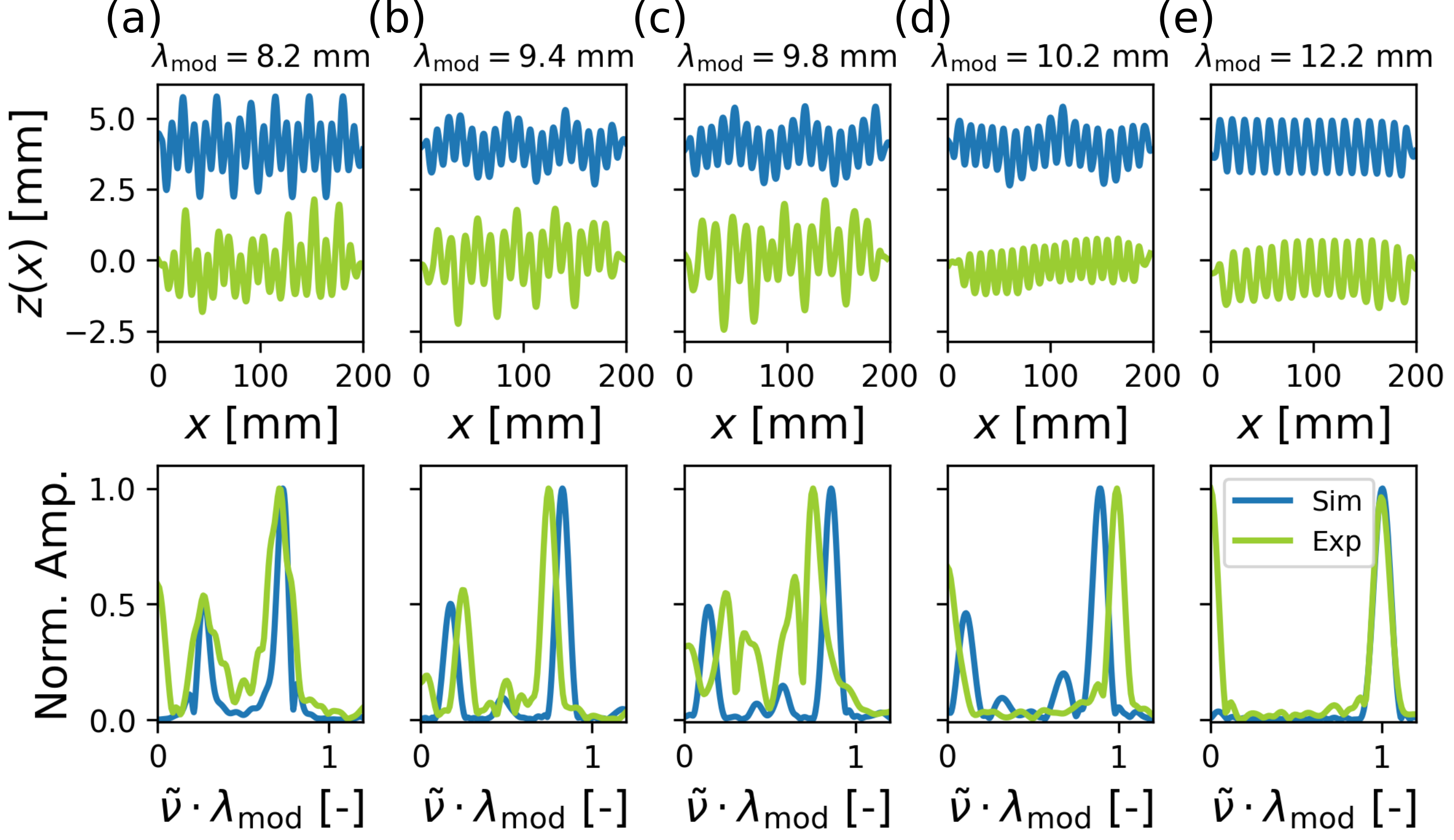}
    \caption{All experimental samples (green) compared with post-buckling simulations (blue) for $\lammod = 8.2$, 9.4, 9.8, 10.2, and 12.2 mm.}
    \label{fig:si_exp_vs_sims_all}
\end{figure}

All our experimental results can be seen in Fig.~\ref{fig:si_exp_vs_sims_all}. 
We find strong agreement for $\lammod = 8.2$ mm and 12.2 mm, and moderate agreement for all other samples.

\vspace{5 mm}
\section*{Data availability statement}
The data that support the findings of this study are openly available in [Github] at \url{https://github.com/bertoldi-collab/wavenumber-locking}

\end{document}